# The frequency of window damage caused by bolide airbursts: a quarter century case study

Nayeob Gi[a], Peter Brown[b,c], Michael Aftosmis[d]

[a]*Department of Earth Sciences, University of Western Ontario, London, Ontario, Canada N6A 3K7 (ngi@uwo.ca)*
[b]*Department of Physics and Astronomy, University of Western Ontario, London, Ontario, Canada N6A 3K7 (pbrown@uwo.ca)*
[c]*Centre for Planetary Science and Exploration, University of Western Ontario, London, Ontario, Canada N6A 5B7 (pbrown@uwo.ca)*
[d]*Computational Aerosciences, NASA Advanced Supercomputing Division, NASA Ames Research Center, Moffett Field, CA, 94035, USA (Michael.Aftosmis@nasa.gov)*

**Abstract**

We have empirically estimated how often fireball shocks produce overpressure ($\Delta P$) at the ground sufficient to damage windows. Our study used a numerical entry model to estimate the energy deposition and shock production for a suite of 23 energetic fireballs reported by US Government sensors over the last quarter century. For each of these events we estimated the peak $\Delta P$ on the ground and the ground area above $\Delta P$ thresholds of 200 and 500 Pa where light and heavy window damage, respectively, is expected. Our results suggest that at the highest $\Delta P$ it is the rare, large fireballs (such as the Chelyabinsk fireball) which dominate the long-term areal ground footprints for heavy window damage. The height at the fireball peak brightness and the fireball entry angle contribute to the variance in ground $\Delta P$, with lower heights and shallower angles producing larger ground footprints and more potential damage. The effective threshold energy for fireballs to produce heavy window damage is ∼5 - 10 kT; such fireballs occur globally once every one to two years. These largest annual bolide events, should they occur over a major urban centre with large numbers of windows, can be expected to produce economically significant window damage. However, the mean frequency of heavy window damage ($\Delta P$ above 500 Pa) from fireball shock waves occurring over urban areas is estimated to be approximately once every 5000 years. Light window damage ($\Delta P$ above 200 Pa) is expected every

## INTRODUCTION

Understanding the small (1 - 20 m) near-Earth objects (NEOs) population has become more important in recent years as the damage risk from these objects appears to be greater than previously thought (Chapman and Morrison, 1994; Brown et al., 2013). The estimated flux of small impactors suggests that a 1 m diameter object strikes Earth every 1-2 weeks, a 10 m object every 15-20 years while a 20 m diameter NEO is expected to collide with the Earth every 50-100 years (Brown et al., 2002; Boslough et al., 2015; Harris and D'Abramo, 2015). For these small objects the atmosphere usually absorbs the majority of the initial energy and a ground-level airburst is avoided. In this size range, the ground damage caused by a bolide is most likely to be due to the airburst shock wave (Chapman and Morrison, 1994; Hills and Goda, 1998; Collins et al., 2005; Rumpf et al., 2017; Collins et al., 2017), which can result in a surface airblast sufficient to cause property damage and/or loss of life, should it occur over a populated area (Boslough and Crawford, 2008).

The February 15, 2013 airburst proximal to the city of Chelyabinsk in Russia, was the first recorded impact producing an air blast leading to widespread window damage (Brown et al., 2013) in an urban area. The shock wave impacting the city caused in excess of $60 M in damage, mostly through breakage or cracking of windows (Popova et al., 2013).

As demonstrated by the Chelyabinsk event, at the lowest threshold where impactors are expected to just barely cause air blast damage at the ground, window breakage is the most likely damage modality. As the size-frequency distribution of impactors is a power-law, these are also the most likely events to occur. This problem is similar to the sonic boom threshold damage problem encountered in aeronautics (Clarkson and Mayes, 1972; Seaman, 1967). Prior to Chelyabinsk, however, studies of air blast damage from airbursts have focused on the ground footprint under the airburst where $\Delta P$ is very large. These works most often use the Hills and Goda (1998) criteria of the ground footprint where the $\Delta P$ exceeds 28 kPa (e.g. Collins et al. (2005), Toon et al. (1997)), which is an overpressure at which trees are toppled and buildings seriously damaged.

The goal of our study is to quantify the expected incidence of window breakage from the ground level shocks (air blasts) produced by fireballs (airbursts). As discussed later, addressing this problem primarily requires knowledge of the height and magnitude of the energy deposition profile for an airburst.

There are two approaches to addressing this question.

The first, is to model in detail the ablation, fragmentation and subsequent energy deposition of a hypothetical meteoroid and then propagate the resulting shock to the ground. This approach has been widely used employing both analytical models (Chyba et al., 1993; Hills and Goda, 1993, 1998; Collins et al., 2005) and numerical hydrocodes (Boslough and Crawford, 1997; Shuvalov and Trubetskaya, 2007). Recently, Register et al. (2017) has bridged such analytical models combining them with elements of high fidelity strength-based models in hydrocodes. In addition, very high fidelity numerical entry models using detailed estimates of meteoroid strength and shock behaviour (Avramenko et al., 2014; Shuvalov et al., 2013; Register et al., 2017; Robertson and Mathias, 2017; Collins et al., 2017), have been validated against the observed ground-level $\Delta P$ from Chelyabinsk (Aftosmis et al., 2016). Extending this analysis to a large population of hypothetical impactors could produce a statistical estimate of overpressure footprints on the ground as a function of time.

Recently, Mathias et al. (2017) has merged modern entry models and blast models to produce a comprehensive global asteroid impact risk assessment incorporating all damage modalities using a Monte Carlo approach, while Collins et al. (2017) has done a similar analysis focused on blast wave damage alone. The advantage of these approaches is the ability to perform large numbers of realizations exploring wide swaths of parameter space to fully characterize damage modalities, limited only by the underlying physical assumptions of the numerical entry models.

A drawback of these "physics-first" approaches is the need to assume the properties and response to atmospheric entry of hypothetical meteoroids, notably strength and fragmentation behaviour, together with parameters which may require tuning and which subsequently drives the resulting energy deposition profile.

A second approach to estimating the energy deposition profile is to rely on empirical relationships to bound the solution space for a set of real world-cases. This approach becomes particularly useful if we have airbursts for which some information is available (such as energy, speed and height at peak brightness). In such cases, we can reconstruct the energy deposition profile using empirical estimates of peak brightness as a function of total energy and strength when coupled to a numerical entry model. Fortunately, such a dataset of fireballs has recently become available.

In this study, we adopt the second approach to present an empirically-focused analysis of how often fireballs may be expected to produce $\Delta P$s at the ground sufficient to damage windows. We do this by simulating in detail a set of energetic fireballs ($E > 2\,\text{kT}$) reported on the NASA Jet Propulsion Laboratory (JPL) fireball webpage (NASA Jet Propulsion Laboratory, 2018). These data consist of over 600 bright fireballs recorded by US Government sensors in the last 25 years. This data is collected by US Government sensors which monitor Earth's surface and atmosphere for events of interest, and is provided to NASA for scientific study of natural objects impacting the Earth.

The specific fireballs chosen for our analysis can be found in Supplementary Material (SM) section A Table S1. To be included in our dataset, an estimate of total fireball energy (which must be $> 2\,\text{kT}$), velocity and height at peak brightness in addition to location must be reported. The JPL website does not explicitly report energy deposition as a function of height.

To estimate energy deposition as a function of height we will make use of a Monte Carlo numerical approach based on application of an analytic entry model, namely the Triggered Progressive Fragmentation Model (TPFM) of ReVelle (2005). Our aim is to reproduce as accurately as possible the maximum energy deposition of each event, where we expect most of the damaging shock to originate. Using these estimates of the energy deposition and its probable range for a given fireball, we couple the output of the TPFM model with an analytic weak shock model (ReVelle, 1976) to estimate the $\Delta P$ footprint on the ground.

As our dataset consists only of 25 years of fireball measurements as reported on the JPL webpage, we are limited by small number statistics. Though beyond the scope of the current study, it would be useful to simulate an even larger set of fireballs using a full Monte Carlo simulation approach following the procedure of Mathias et al. (2017) for comparison to our results.

In our approach, the energy deposition profiles away from the location of the peak energy deposition height are expected to be less accurately reproduced, but we anticipate this will not change the $\Delta P$ computed very much. For most fireball events, we do not have enough data (particularly observed light curves) to validate whether our generic TPFM approach produces reasonable energy deposition profiles. To check this assumption and to validate our approach of generating model energy deposition profiles from empirical relations, we will apply our generic approach to five well-constrained fireballs and attempt to demonstrate the goodness of fit between our model results and the observations. These five fireballs are found among the JPL data, but in addition to the data given by that source other publications provide known trajectories, and (most importantly) observed light curves. These light curves are an indirect measurement of the fireball's associated energy deposition.

These calibration fireball events are:

1. Feb 1, 1994 - the Marshall Islands fireball (Tagliaferri et al., 1995)
2. Jan 18, 2000 - the Tagish Lake fireball (Hildebrand et al., 2006; Brown et al., 2002)
3. Mar 27, 2003 - the Park Forest fireball (Brown et al., 2004)
4. Sep 3, 2004 - the Antarctica fireball (Klekociuk et al., 2005)
5. Jul 23, 2008 - the Tajikistan superbolide (Konovalova et al., 2013)

For these five cases we can independently check our model energy deposition profiles against the observed energy deposition (light) curve.

Similarly, the analytic weak shock model has not been previously validated. Here we report explicit validation of the weak shock model for estimating fireball overpressure using two approaches. First, we compare the model predictions to a known ground-truthed event (the Genesis sample-return entry, (ReVelle, 2005)) where overpressure at the ground was measured. Secondly, we compare weak shock overpressure predictions for two of our calibration fireball events with well measured light curves to predictions from the extensively validated and higher fidelity Cart3D model, described by Aftosmis et al. (2016).

For all fireball events in our study we have computed the area at ground-level where the $\Delta P$ is large enough to break windows. From this suite of $\Delta P$-Area per unit time estimates, we then estimate the frequency with which we expect fireballs to produce window damage over urban areas on a global scale, assuming this 25 year interval is representative of the cumulative overpressure footprint from fireballs in any 25 year period.

**BACKGROUND**

**Window breakage - general considerations**

Window breakage is a significant damage mode in airblasts (Glasstone and Dolan, 1977). Injuries are commonly due to flying glass. In general, structural damage from airblasts is largely determined by the duration and amplitude of the blast wave (Needham, 2010). However, small and light structural elements, such as windows, require only a short period of vibration (up to ∼0.05 sec) and small plastic deformation to break. Therefore, the breakage of window glass is mostly determined by peak $\Delta P$, the maximum pressure caused by a blast wave above the ambient atmospheric pressure, without significant considerations needed for the duration of the blast wave (Glasstone and Dolan, 1977; Pritchard, 1981).

Window breakage is a complex process (Zhang and Hao, 2016). For a given shock geometry, $\Delta P$ and pulse duration, window failure depends on factors such as window thickness, area, method of attachment to frame, defect/microcrack density and damage history (Pritchard, 1981). Identically produced and mounted windows will not fail under the same conditions, because of microstructural variability (Hershey and Higgins, 1976). As a result, window breakage by airblasts is treated statistically with prediction models using empirical relations scaled to window thickness and area with various simplifications (Fletcher et al., 1980). In particular, the pulse duration/impulse is a critical factor that can influence window damage levels. The $\Delta P$-impulse diagram given by Gilbert (1994), however, shows that at high charge weights ($>0.02\,\text{kT}$), $\Delta P$ is the only factor that determines structural damage. At low charge weights ($<5\times10^{-4}\,\text{kT}$), impulse is solely responsible for causing damage to structures. Between these two extremes, both $\Delta P$ and impulse need to be considered to estimate damage levels. However, as most of our fireball sources are comparatively large equivalent charges (on the order of kilotons of TNT), we will assume $\Delta P$ is the only feature of the airblast which needs to be considered in window damage. This is consistent with most past empirical studies of window breakage from large charges (cf. Reed (1992)). To get a simple estimate of the range of $\Delta P$ of interest, we will use a few empirical studies to bound the $\Delta P$ levels at which window damage may be expected to occur. We caution that the relation of window breakage probability to the $\Delta P$ adopted for our study is therefore simple, but we believe it is instructive to address the threshold level for an airwave produced by a fireball at which damage may occur. It is worth noting in what follows that window damage can occur at lower $\Delta P$ levels if the windows are old or already stressed; similarly, newer windows might survive at much higher $\Delta P$ levels.

**Window breakage criteria**

There have been a number of experimental studies giving quantitative estimates of the peak $\Delta P$ which causes window breakage both generally and as a function of thickness/area. Glasstone and Dolan (1977) provide a widely cited approximate $\Delta P$ range of 3.5 - 7 kPa for typical residential large and small glass window failure based on air blasts produced during nuclear tests. Clancey (1972) suggested that the peak $\Delta P$ for small window breakage to be 0.7 kPa while Kinney and Graham (1985) gave the range of typical window glass breakage as 1 - 1.5 kPa. Previous nuclear tests had shown that windows start to break at an $\Delta P$ of about 0.4 kPa, and this is the standard adopted in ANSI (1983).

However, a fundamental problem with these earlier studies is the lack of consideration for the size or thickness of windows. Fletcher et al. (1980) suggest a 50% probability of failure for most face-on windows lies between 0.6 - 6 kPa, showing explicit dependence on window area based on the experimental results of Iverson (1968).

In exploring all the literature on window breakage, we found one study in particular which used real-world data, explicitly included window sizes and was consistent with other studies. In this work, Reed (1992) derived empirical relations for predicting airblast damage to windows based on records of window breakage due to the 1963 Medina facility explosion, an accidental explosion of 50 tonnes of chemical high explosives near San Antonio, Texas. Reed (1992) explored the relationship between window breakage probability and incident $\Delta P$ for typical San Antonio window panes, which are taken to be a single-strength glass, 0.6 m × 0.6 m × 2 mm thick. Gilbert (1994) derived a probit equation from the Reed (1992) relationship, namely:

$$Y = -4.77 + 1.09 \, ln\left(p_e^\circ\right) \qquad (1)$$

where $Y$ is the probit and $p_e^\circ$ is the peak effective $\Delta P$ (Pa) experienced by Reed's standard pane. We take the peak incident $\Delta P$ that would be required for other windows to break from the following equation of Gilbert (1994):

$$p^\circ = \frac{\left(\frac{A/0.372}{t/0.002}\right)}{p_e^\circ} \quad (2)$$

where $p^\circ$ is the peak incident $\Delta P$ (Pa), $A$ is the pane area (m$^2$), and $t$ is the glass thickness (m). Using Eq. 1 and 2, Fig. 1 shows the window breakage probability as a function of incident $\Delta P$ for typical window sizes in urban areas following Gilbert (1994). We note that our range and breakage probability are broadly consistent with earlier studies, in particular it is comparable to changes in $\Delta P$ as a function of area values summarized in Fletcher et al. (1980).

**Data for window breakage from the Chelyabinsk airburst and adopted criteria**

The February 15, 2013 Chelyabinsk airburst is the only fireball for which widespread window damage was recorded. One challenge with estimating window breakage percentages was the rapid replacement of windows after the event due to the winter conditions at the time.

Brown et al. (2013) used videos from the time of the event or immediately (1-2 days) afterward to attempt to quantify window damage and therefore remove any window replacement bias. They examined a total of 5415 windows in Chelyabinsk visible in videos with known geolocation. They categorized windows into four area groupings: A: 0-0.5 m$^2$, B: 0.5-1 m$^2$, C: 1-1.5 m$^2$, and D > 1.5 m$^2$. The majority of windows fell in categories B and C: 1810 (33%) windows being Category B and 2258 (42%) windows being Category C, corresponding to the shaded area in Fig. 1. The average percentage of standard window breakage based on Eq. 1 and 2 is expected to be ∼0.01-0.7% at 0.2 kPa, ∼0.4-7% at 0.5 kPa, and ∼25-60% at 3 kPa, the latter range being consistent with the weighted average of 20% breakage reported in Brown et al. (2013) for class B and C windows.

There was a strong variability across the city in window breakage, with some sections in the northern part of the city experiencing much larger breakage percentage, suggesting that local values may deviate by up to a factor of two from the nominally reported value near 3 kPa.

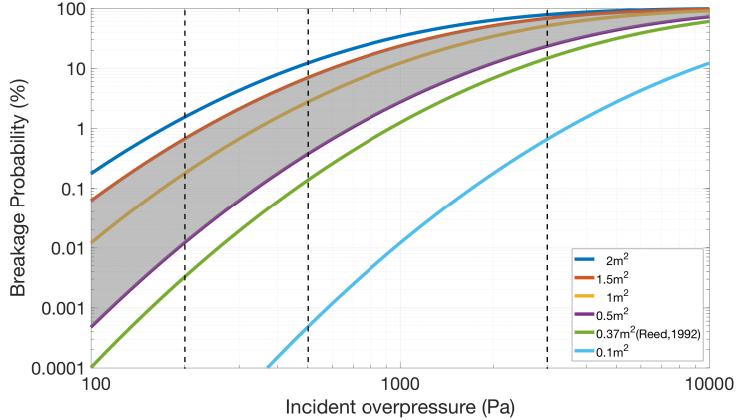

Figure 1: Window breakage probability as a function of incident $\Delta P$ for six typical window sizes in urban areas. Colored lines represent different window pane areas. The green line corresponds to the Reed (1992) single-strength glass (0.6 m × 0.6 m × 2 mm thick). The shaded region includes sizes representative of those found in most urban areas. Dashed vertical lines indicate reference incident $\Delta P$s of 0.2 kPa, 0.5 kPa and 3 kPa. Note that based on the work of Fletcher et al. (1980) increasing the thickness from 2 mm to 6 mm increases the corresponding breakage $\Delta P$ curves by a factor of four.

Independent estimates of the $\Delta P$ in Chelyabinsk are available from several sources. Brown et al. (2013) used the measured velocity of glass shards from several videos and empirical relations of $\Delta P$ versus expected shard speed to estimate a $\Delta P$ of 2.6 kPa. Avramenko et al. (2014) measured the apparent dynamic pressure of the air blast by the observed jump in lateral velocity of car exhaust in two videos to estimate an equivalent $\Delta P$ of 1.6 - 1.9 kPa.

Comparing these estimates to those obtained from our empirical window breakage relations (e.g. Fig. 1), we see a better than factor of two agreement. Given the variability in $\Delta P$ expected in an urban area due to reflections, caustics and large scale shock interference, this is remarkably consistent. We suggest that this confirms the basic validity of our adopted empirical relations.

As such, we use Fig. 1 as our baseline estimate to quantify window breakage.

We will examine the areal footprint on the ground under our modelled airbursts where $\Delta P$s exceed $200\,\text{Pa}$ and $500\,\text{Pa}$, denoting these hereafter as $\Delta P(200)$ and $\Delta P(500)$ and describe them as light and heavy window damage thresholds respectively.

These two $\Delta P$ thresholds correspond approximately to the levels at which large windows ($2\,\text{m}^2$) have a 1.5% and 12% breakage probability respectively. Similarly, standard urban windows (with $0.5 < \text{A} < 1.5\,\text{m}^2$) would have a 0.01- 0.7% and 0.4- 7% probability of breakage for $\Delta P(200)$ and $\Delta P(500)$. In practical terms, these breakage probabilities bracket the ranges at which window damage from sonic booms are cited as producing damage claims in urban areas (Clarkson and Mayes, 1972).

## ANALYSIS METHODOLOGY AND MODEL VALIDATION

### Triggered Progressive Fragmentation Model (TPFM)

To estimate ground-level $\Delta P$, we must first estimate the energy deposition as a function of height for each fireball. Following ReVelle (2005) we use the analytic Triggered Progressive Fragmentation Model (TPFM), which allows explicit inclusion of a simple fragmentation model once a body's tensile strength is exceeded to simulate energy deposition and ablation. Our approach attempts to best match the peak energy deposition; heights above and below this point are expected to have poor (factor of several) discrepancies in modeled versus observed energy deposition.

The model is based on analytically solving coupled differential equations for the meteoroid speed (Eq. 3a) and mass (Eq. 3b) to determine the height of the meteoroid as a function of its speed:

$$\frac{dv}{dt} = -\frac{\rho_{atm} C_D A v^2}{2m} \tag{3a}$$

$$\frac{dm}{dt} = -\frac{\rho_{atm} C_H A v^3}{2Q} \tag{3b}$$

where $v$ is the meteoroid speed (km/s), $m$ is the mass (kg), $t$ is the time (s), $\rho_{atm}$ is the atmospheric density (kg/m$^3$), $C_D$ is the drag coefficient, $C_H$ is the heat transfer coefficient, $A$ is the cross sectional area (m$^2$), and $Q$ is the meteoroid heat of ablation (ReVelle, 2005).

The model allows the ablation coefficient ($\sigma = \frac{C_H}{2QC_D}$) to change through variable drag, heat transfer coefficient and heat of ablation with height according to the flow regime, speed and material properties as described in ReVelle (1979) and modifications to that original approach outlined in ReVelle (2005). The atmosphere is non-isothermal with the atmospheric mass density profile taken from the NRLMSIS-00 model of Picone et al. (2002) for the location and time of each simulated event.

Fragmentation for each simulation realization is randomly permitted to generate 0 to 1024 fragments in total, with each fragmentation episode doubling the number of fragments. In this manner, each time the dynamical pressure exceeds the meteoroid strength (specified in the simulation according to empirical criteria - see section Empirical constraints for TPFM model), the meteoroid splits in half and the new fragment is assumed to ablate without any shielding effects from the leading fragment as described in ReVelle (2005).

Ablation is assumed to occur such that all fragments remain as spheres. Each succeeding fragment generation is assumed to have a higher strength, with the strength increment based on a Weibull distribution with the Weibull scaling factor exponent $\alpha$ chosen in the simulation randomly between values of 0.2 - 0.5 (Popova et al., 2011). Specifically, for each simulated run, each succeeding fragmentation generation $i$ has a strength $S_i$ compared to the previous fragment strength $S_{i-1}$ of Popova et al. (2011):

$$S_i = S_{i-1} \left( \frac{m_{i-1}}{m_i} \right)^\alpha \tag{4}$$

where $\alpha$ is chosen at the start of the run and the same exponent used throughout that particular run.

Each simulation ceases when less than 1% of the original kinetic energy of the fireball remains. The resulting TPFM energy deposition per unit path length output is then coupled with the ReVelle (1976) weak shock model to estimate $\Delta P$ on the ground.

The starting data used for our simulations is from the JPL fireball webpage which provides basic information on hundreds of real fireballs including in some cases entry angle, entry speed, height at the peak brightness, and total impact energy.

TPFM as a bolide ablation model computes mass loss, light production, and fragmentation associated with the atmospheric entry of fireballs. In general, the model input parameters are tuned to match the observed light curves and dynamics of fireballs. In these cases the TPFM fits may then provide estimates of the initial meteoroid properties including mass, porosity, strength, ablation rate and fragmentation behaviour. This forward modelling application of TPFM has already been applied to a number of past events (ReVelle, 2005, 2007; Brown et al., 2013).

In our study, we use each TPFM run as a single realization to try and match the available data from the JPL site of a particular fireball. For each fireball being simulated a number of input parameters are approximately known from JPL data (e.g. initial speed (km/s), entry angle (deg), initial energy (kT)). Other parameters which are not known a priori are randomly chosen from broader distributions in a Monte Carlo sense (e.g. porosity, strength, increment in fragment strength from the Weibull distribution).

The simulations allow the fireball energy to vary by up to a factor of two compared to the JPL-reported value following the theoretical arguments about variation in luminous efficiency in Nemtchinov et al. (1997). This distribution was assumed to be uniform. Based on this range of kinetic energy and our known velocity, the corresponding mass range was then computed. The simulations had porosity variations from 0-95% (uniformly and randomly distributed) to cover all possible types of meteoroids, except for the Tagish Lake meteorite/fireball, where we used 40-95% porosity, the lower limit determined from the recovered meteorites (Hildebrand et al., 2006). The initial meteoroid radius was computed using an assumed grain density of $3500\,\mathrm{kg/m^3}$ together with the previously chosen porosity and mass as estimated from the Monte Carlo energy and known velocity. These initial parameters were then used as inputs to the TPFM model.

The full range explored for these input parameters among our five calibration fireball events are shown in Table S2 in the Supplementary Material (SM) section B.

For each of the five calibration fireballs (i.e. those having measured light curves in addition to all metric data), we simulated 1000 runs based on the initial parameters given in Table S2 in the SM section B. We then down selected to only model runs that match our observational constraints (which are described in detail in the next section): namely simulations which have peak energy deposition within 3 km of the observed height at peak brightness (Fig. S1(a) in the SM section B), are within factor of two of the total reported energy (Fig. S1(b) in the SM section B) and show a peak magnitude which correlated with the total energy consistent with the population of bolides as a whole as shown in Fig. 3. The resulting range of explored parameters is summarized in Table S3 in the SM section B.

A similar process was used for the remaining 18 fireballs we later simulated, except, of course, in those cases no light curves are available. For these cases a suite of 10,000 realizations was created. From this broad suite of model runs we chose a subset which also match empirical relations (see section Empirical constraints for TPFM model) and compute the corresponding range of energy deposition to estimate median and maximum $\Delta P$ fields at the ground.

**Empirical constraints for TPFM model**

To select among our 10,000 simulations those which are most probable on physical grounds, we develop some empirical constraints from the population of bright bolides as a whole as a filter to select the most appropriate model runs. The first constraint is provided by the observational measurement of the height at peak brightness published on the JPL website. The height at peak brightness is known to have an accuracy of order 3 km from an earlier study where several JPL fireballs also observed from the ground were compared in detail (Brown et al., 2016). From that work, the measured height of peak brightness as a function of velocity for meter-sized impactors was determined as shown in Fig. 2a. This is equivalent to an estimate of the strength (calculated as the dynamic pressure for each object at its fragmentation height) as shown in Fig. 2b. The initial fragmentation occurs earlier than the point of peak brightness, so using the latter height provides an upper limit to the strength of the meteoroid (cf. Collins et al. (2017)). Our estimation neglects deceleration and in general, deceleration is negligible for most of these fireballs as they are such large objects (multi-meter-sized).

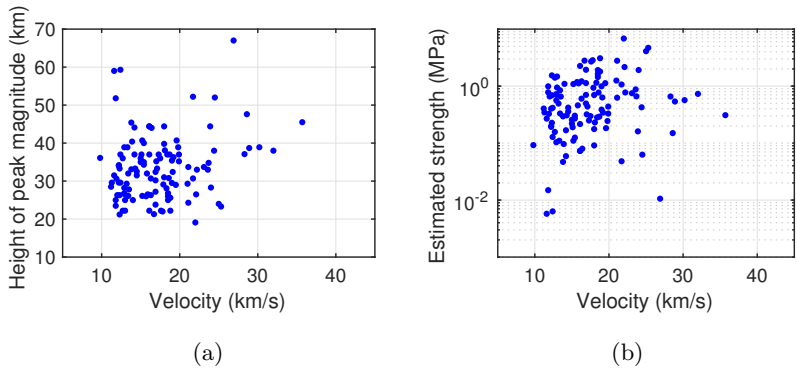

Figure 2: The distribution of (a) measured height of peak brightness and (b) the estimated strength based on this height (making the strength an upper limit to the global strength of the meteoroid) as a function of initial velocity for all meter-size objects (Brown et al., 2016).

From TPFM modelling we find that in practice we can match the height of peak brightness assuming the first fragmentation begins between one and two atmospheric scale heights above the height of peak brightness, depending on the number of assumed fragmentation episodes. While the global strength can be roughly matched in this manner, the fragmentation behaviour is still unspecified. We expect the height of peak brightness to correlate with the total energy of an event. The vertical spread in this correlation is a proxy for the degree of fragmentation.

Hence, to constrain the simulations further, we use the relationship between observed peak magnitude (radiated power) and total impact energy derived from the dataset reported in Brown et al. (2002) as given in Brown (2016) and shown in Fig. 3. We require each realization to fall within the 2σ prediction intervals about the regression of Fig. 3. The best-fit regression to larger events (E > 1 kT) is given by

$$M_{peak} = -21.2 \pm 0.1 - (2.30 \pm 0.16) log E \quad (5)$$

where $M_{peak}$ is the peak brightness and $E$ is the total energy (kT). This peak brightness-energy filtering selects for model runs which have fragmentation behaviour physically similar to the meter-sized impactor population as a whole. We expect some deviation for very weak objects or objects entering at unusually shallow angles.

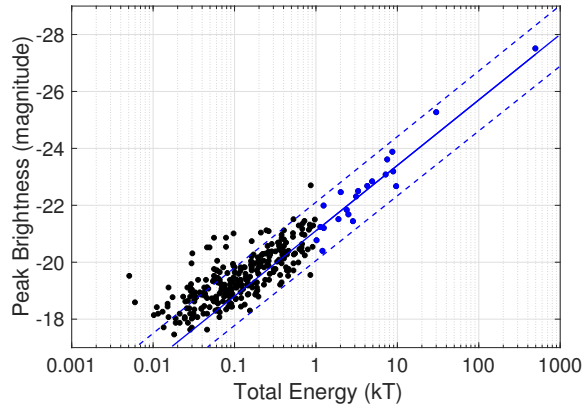

Figure 3: The distribution of measured peak brightness as a function of total impact energy (Brown, 2016). The dataset consists of 300 optical events from Brown et al. (2002) and the Chelyabinsk event (∼500 kT) from Brown et al. (2013). We see some deviation in the trend for smaller events (E < 1 kT). The blue solid line is a direct linear regression fit for events with E > 1 kT also shown as blue circles. The blue dashed line shows the 2σ prediction interval about the regression.

Finally, we also filter the model runs by requiring that the simulated total energy is within a factor of two of the JPL reported total impact energy based on modelling of the luminous efficiency which shows a similar variation (Nemtchinov et al., 1997).

An example of the resulting model plots of filtered runs (i.e. those which produce maxima within 3 km of the reported height at peak brightness, lie within the $2\sigma$ prediction interval of Fig. 3, and span a factor of two compared to the JPL reported energy) are shown in the SM section B Fig. S1.

Our focus is to generate model runs that reproduce the main portion of the observed energy deposition curve. As we show later in the section on empirical ablation modelling: Five calibration case studies and the SM section E, we obtain reasonable agreement between our maximum Monte Carlo TPFM energy deposition profiles and the observed profiles for five calibration fireball events. On this basis, we do not explore the full parameter space, but limit to just those runs which match our empirical criteria. For the additional 18 fireball events we simulate, we have no data on detailed energy deposition, so detailed examination of all possible free parameters which might fit the sparse data is not feasible.

This Monte Carlo simulation procedure is followed using the TPFM code one thousand times for each of the five calibration events and ten thousand times for each of the additional 18 JPL fireballs in our study. These final filtered runs for each event provide the estimated range of energy deposition values as a function of height that in turn form the basis of the input for the next step in the simulations; namely estimating the $\Delta P$ at the ground.

**ReVelle weak shock model**

The intense energy deposition produced along the bolide trail mimicks a strong cylindrical line shock near the trail, decaying to a weak-shock and eventually to a linear acoustic wave (Edwards, 2009). This cylindrical shock propagates perpendicular to the meteoroid trajectory. To numerically map the footprint of the $\Delta P$ at the surface, we simulated a grid of points at the ground that follows this specular geometry (Fig. 4). These points were computed every 0.01 degrees in latitude and longitude at each 1 km increment in height along the fireball trajectory. Fig. 4 shows a limited number of such receiver points for ease of visualization. At each point on the ground, we compute the largest $\Delta P$ and median $\Delta P$ among all accepted TPFM simulation runs and refer to these as maximum $\Delta P$ and median $\Delta P$. From the array of maximum and median $\Delta P$, we find the single point having the largest overpressure on the ground; we refer to these as a peak maximum $\Delta P$ and a peak median $\Delta P$, respectively, throughout the paper.

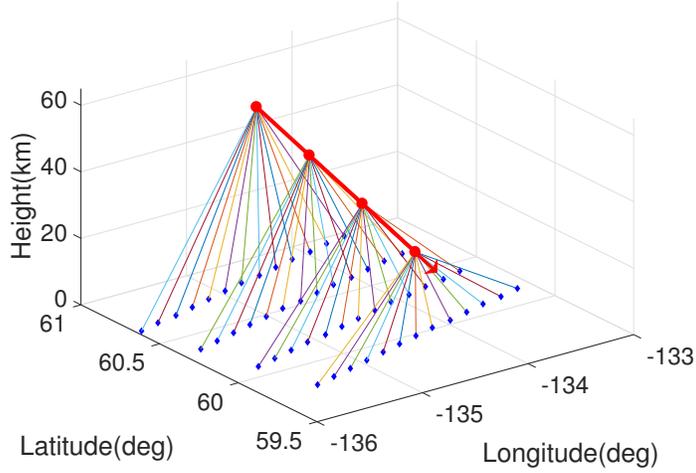

Figure 4: An example showing simulated weak-shock waves reaching a grid of receivers at the ground for the Tagish Lake fireball. The diagram has been simplified for better visualization. Red arrow is the bolide trajectory where red circles show 10 km interval height. Blue diamonds are the receiver points separated by 0.2 degrees.

To compute the expected $\Delta P$ at each receiver point, we adapted the ReVelle (1974, 1976) weak shock model to predict the ground $\Delta P$, using as input the energy deposition model outputs from the TPFM model for each fireball in our study. The ReVelle weak shock model was developed following earlier work on cylindrical shock waves (Sakurai, 1964; Jones et al., 1968; Few, 1969; Tsikulin, 1970). It is an analytical model that requires knowledge of the energy deposition per unit trail length for the bolide and a known geometry between the trail and a receiver point on the ground to estimate ground $\Delta P$. The model makes the following assumptions (ReVelle, 1974):

1. There is no significant deceleration.

2. The trajectory is a straight line, therefore gravitational effects are negligible.

3. Only rays that propagate downward and are direct arrivals at ranges from the source comparable to or smaller than the total fireball trail length are considered.

4. The meteoroid moves much faster (Mach number $> 15$) than the ambient sound speed, so the energy release is approximately instantaneous across the entire trail.

5. The Knudsen number is $< 0.05$ so that the body is in continuum flow.

6. Shock reflections and interference are ignored.

Note that one consequence of assumption 3 is that we expect the weak shock model to be less accurate for very steep entry angles. Note also that some of the numerical entry models including Popova et al. (2013) suggest that shock reflection and interference may produce significant changes to the ground overpressures and we are ignoring this effect.

According to line source blast wave theory, the blast radius, ($R_o$) at any point along the trail, is defined as (Tsikulin, 1970):

$$R_o = \left(\frac{E_o}{P_o}\right)^{\frac{1}{2}} \tag{6}$$

where $E_o$ is the total energy per unit trail length and $P_o$ is the ambient hydrostatic atmospheric pressure at the source height. Physically, the blast radius is the distance from the center of the meteoroid trail to where the shock $\Delta P$ drops to roughly the ambient background pressure. It corresponds to the distance away from the meteoroid trajectory where the expansion work done by the shock to move the surrounding atmosphere equals the deposited explosion energy (Few, 1969). We calculate the blast radius ($R_o$) using this fundamental definition in terms of energy deposition per unit trail length. Then, $R_o$ is used as an input for the weak shock model to determine the predicted overpressure (pressure caused by a shock wave) on the ground to gauge blast damage. According to the weak shock model, the shock wave reaches its fundamental period $\tau_o$ after travelling approximately ten times the blast radius. We may apply the ReVelle (1976) weak shock model to calculate the fundamental period by inverting the fundamental frequency of the wave, which is given by:

$$\tau_o = \frac{1}{f_o} = \frac{2.81 R_o}{C_s}, \tag{7}$$

where $R_o$ is the blast radius and $C_s$ is the speed of sound. Beyond $R_o$, the shock propagates as a weak nonlinear wave. We assume that the shock propagates to the ground as a weak-shock and does not undergo a transition to linearity, in contrast to the original ReVelle (1974) theory which always assumes such a transition. Physically, this has been shown to be a good approximation as described by Silber et al. (2015) and is appropriate to our short ranges for the large energy fireballs of our case study. Note that if we assume transition to linearity our estimated $\Delta P$ would increase in all cases, so this assumption makes our $\Delta P$ conservative. Details of the algorithms can be found in ReVelle (1974, 1976), Edwards (2009), and Silber et al. (2015).

**Validation of the ReVelle Weak Shock Model**

While having been used extensively in the literature for several decades, the methodology proposed originally in ReVelle (1974) has not been validated.

The most desirable validation would be to compare measured overpressure on the ground for a fireball which also has a measured energy deposition profile. Unfortunately, Chelyabinsk is the only such event to date. As discussed later, the blast radius for Chelyabinsk is so large that except at larger slant ranges (outside the city of Chelyabinsk) the model is largely inapplicable.

*Genesis re-entry*

The first approach we used to check the validity of the weak shock model was to compare the ground overpressure computed from the model against observed ground-based infrasound amplitude. Here we validate using the re-entry of the Genesis sample return capsule which occurred on September 8, 2004. Its entry speed of 11.0 km/s with an entry angle of 8° from the horizontal (Jenniskens et al., 2006) is within the range of applicability of the weak shock formalism. An infrasound signal was detected at a portable microphone array located at Wendover, Nevada. ReVelle et al. (2005) analyzed infrasound signals arriving from Genesis using the InfraTool component of the wave processing software package Matseis (Harris and Young, 1997; Young et al., 2002). The measured maximum amplitude was 3.995±0.1585 Pa (ReVelle et al., 2005). Using the results from ReVelle et al. (2005) the source height at this ground location was 41 km and the blast radius from the measured deceleration of the capsule was 16 m.

Applying the weak shock model and assuming no transition to a linear wave produces an estimate of 4.4 Pa for the overpressure, while inclusion of a transition to a linear wave results in a model estimate of 7.9 Pa. Throughout this work, we have assumed no transition to a linear wave as this is both most consistent with available observations for smaller energy meteors (Silber et al., 2015) and gives a conservative value. This result suggests that the nominal weak shock overpressure predicted by the model is very close to observations in this case while the overpressure including a transition to a linear wave is too high by a factor of two.

*Comparison to results from Cart3D*

A second approach to validation of the weak shock model is comparison with the shock/overpressure predictions from Cart3D, a fully conservative, finite volume solver which uses a multilevel Cartesian mesh. In our application, Cart3D is employed to propagate in time and space the shock from the meteor source to the ground, using as input an initial energy deposition profile (Aftosmis et al., 2016). This code (Aftosmis and Berger, 1998) has been extensively validated in a number of physical settings, including reproduction of the ground overpressure for Chelyabinsk (Aftosmis et al., 2016).

As it is computationally intensive, we have chosen only two of our five calibration fireballs with measured light curves for this comparative analysis, namely the February 1, 1994 Marshall Islands event and the September 3, 2004 Antarctica event.

Using the same energy deposition profile, the overpressure footprint at the ground for both of these events is shown in Fig. 5 and the areal ground footprints summarized in Table. 1.

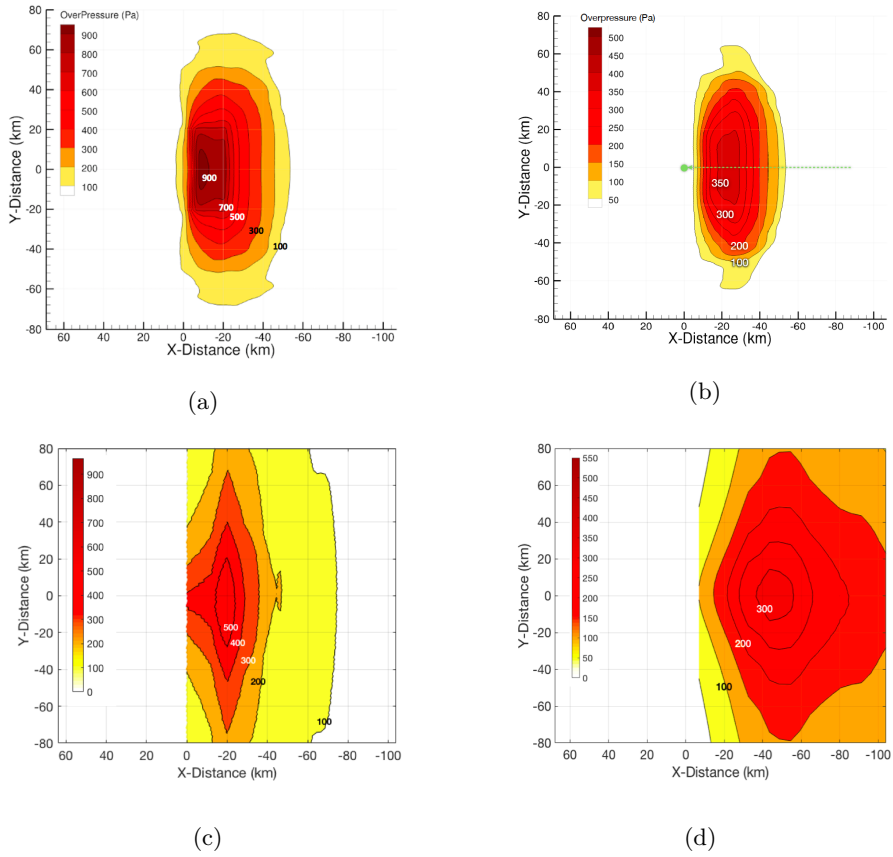

Figure 5: Comparison of the predictions for ground-level maximum weak shock $\Delta P$ (Pa) for the Marshall Islands (left panels) and the Antarctica fireball (right panels) computed by Cart3D (top row) and the weak shock model (bottom row).

We see from comparison for these two events that the weak shock model produces similar results to Cart3D, in particular within a factor of ∼2 for ground area footprint for $\Delta P(200)$ for both fireballs. The Marshall Islands event has a relatively larger difference at $\Delta P(500)$, though the absolute area difference is still very small compared to the much larger footprint of Chelyabinsk.

|  |  | Cart3D | Weak Shock Model | Weak Shock Model (with linear transition) |
|---|---|---|---|---|
| Marshall Islands | Peak $\Delta P$ (Pa) | 937 | 594 | 886 |
|  | $\Delta P(200)$ (km$^2$) | 4784 | 7206 | 25599 |
|  | $\Delta P(500)$ (km$^2$) | 1436 | 259 | 2097 |
| Antarctica | Peak $\Delta P$ (Pa) | 361 | 326 | 539 |
|  | $\Delta P(200)$ (km$^2$) | 1984 | 3920 | 20494 |
|  | $\Delta P(500)$ (km$^2$) | 0 | 0 | 383 |

Table 1: Comparison of peak $\Delta P$ (Pa) and threshold $\Delta P$-areas (km$^2$) computed by Cart3D and the weak shock model as used in our study. The predictions of the weak shock model assuming a transition to linearity as in the original ReVelle (1974) theory is also shown for comparison.

The peak overpressure predicted from Cart3D for the Antarctica event agree to within 10% with the weak shock result, while the Marshall Island fireball prediction is roughly 50% above the weak shock prediction. Interestingly, using the complete weak shock theory including a transition to linearity provides much better agreement for the Marshall Island event with the predictions of Cart3D.

Fig. S3 and S4 in the SM section C shows an overview of the evolving shock using Cart3D simulation for both events. For these events where the energy deposition shows a strong maximum near the end of the trail (a flare), it is clear that the assumption of a cylindrical shock remains valid. This is because the height gradient of the energy deposition is well matched by the atmospheric mass density gradient in these cases. Thus, the weak shock model provides a reasonable estimation of overpressure even when there is a flare at the end of the flight. More details of the Cart3D modeling are given in the SM section C.

**RESULTS**

**Empirical ablation modelling: Five calibration case studies**

Our modelling approach is designed to estimate peak energy deposition for fireballs where only the height of peak brightness, speed, entry angle and total energy are known. Early or late portions of the bolide entry are entirely unconstrained and we do not expect our approach to produce matches in these parts of the trajectory.

However, we first need to demonstrate that the Monte Carlo TPFM approach using our empirical constraints produces peak energy deposition values similar to observations. Validation of our modelling approach uses five well-documented fireball events, for which we have JPL data, trajectories, as well as the complete observed light curves. The data for these five fireballs are summarized in Table 2.

The observed light curve for each case study is equivalent to an energy deposition curve (assuming a luminous efficiency) which can then be compared to the energy deposition curve produced through our simulated TPFM Monte Carlo runs.

We are assuming for simplicity that the energy deposition is proportional to the light curve with a simple fixed scaling factor. In reality, this conversion is expected to be more complicated (cf. Nemtchinov et al. (1997)) as luminous efficiency should depend on height, speed, object radius and the heated volume does not re-radiate the deposited energy immediately.

The light curve data for each bolide and the details of the luminous efficiency conversion used to produce the energy deposition profiles can be found in the SM section D. For each of these calibration fireball events, the TPFM model fit to the observed energy deposition curve and the resulting predicted ground-level maximum $\Delta P$ plot are generated. We briefly describe the first calibration event, the Marshall Islands fireball, and compare our resulting modelled energy deposition profile fits to observations. The full suite of model fits to the observation for each of the other calibration events can be found in the SM section E. The predicted median and standard deviation $\Delta P$ plots for each of the 5 events can be found in the SM section F. As mentioned earlier, maximum $\Delta P$ refers to the largest $\Delta P$ at any ground point computed from the ensemble of all simulations which met our empirical criteria. Similarly, median $\Delta P$ was calculated from the median of all accepted simulation runs. More details of the specific modelling of each of these events is given in the SM section F.

|  | Marshall Islands | Tagish Lake | Park Forest | Antarctica | Tajikistan |
|---|---|---|---|---|---|
| Date | 1994.02.01 | 2000.01.18 | 2003.03.27 | 2004.09.03 | 2008.07.23 |
| Time (UT) | 22:30 | 16:43 | 5:50 | 12:07 | 14:45 |
| Location | Pacific Ocean (2.7, 164.1) | Canada (60.3, -134.6) | US (41.4, -87.7) | Antarctic (-67.7, 18.8) | Tajikistan (38.6, 68.0) |
| JPL Energy (kT) | 30 | 2.4 | 0.41 | 13 | 0.36 |
| Speed (km/s) | 25 | 15.8 | 19.5 | 13 | 14.3 |
| Entry angle (°) | 45 | 17.8 | 29 | 41.9 | 80 |
| Radiant azimuth (°) | 119.6 | 330.7 | 201 | 82.1 | 278.0 |
| Mass(kg) | $4.2 \times 10^5$ | $8.0 \times 10^4$ | $9.0 \times 10^3$ | $6.4 \times 10^5$ | $1.5 \times 10^4$ |
| Radius(m) | 3.1 | 2.3 | 0.9 | 3.5 | 1.0 |
| Density (kg/m$^{-3}$) | 3500 | 1640 | 3400 | 3500 | 3500 |
| Light curve extracted from | Tagliaferri et al., 1995 | Brown et al., 2002(b) | Brown et al., 2004 | Klekociuk et al., 2005 | Konovalova et al., 2013 |

Table 2: Summary of bolide data for five calibration fireball case studies. Time, location, and energy were taken from NASA JPL fireball website. Speed, entry angle, and radiant azimuth are from the given reference from which the light curve was also extracted. The bolide mass was determined using the JPL estimated kinetic energy derived from the integrated luminous power. The meteoroid radius was computed using the volume of a sphere, where we assume a typical mass density for chondritic meteorites as $\rho$=3,500 kg/m$^{-3}$, except for the Tagish Lake and the Park Forest fireballs where the actual bulk density for the recovered meteorites was used. Tagish Lake was classified as a C2 carbonaceous chondrite (Hildebrand et al., 2006) and Park Forest as an L5 chondrite (Brown et al., 2004).

Note that we have no direct measurements of the $\Delta P$ in these cases so cannot extend the validation to actual measured $\Delta P$ for any of these five events. We note a similar procedure was used for the Chelyabinsk fireball (including use of the TPFM model and the ReVelle (1976) weak shock code) and the match was good (Brown et al., 2013) in the centre of Chelyabinsk, though the technique is unable to estimate the largest $\Delta P$ for Chelyabinsk directly beneath the fireball due to the large blast radius.

*The February 1, 1994 Marshall Islands fireball*

This fireball occurred over the South Pacific and penetrated to a comparatively low altitude, reaching peak brightness at a height of 21 km. It is among the four largest energy events recorded by US Government sensors in the last 25 years. For this first event, the TPFM model runs match well with the observed energy deposition curve as shown in Fig. 6. The observed energy deposition curve depicts two peaks as the meteoroid undergoes explosive disintegration at heights of 34 km and 21 km. The disintegration at 34 km is not well reproduced with the model runs, however the major disintegration at 21 km is in good agreement with the model. The peak brightness occurs where the peak energy deposition occurs. We assume that the peak $\Delta P$ is dominated by the peak energy deposition. Therefore, in this run as with all our other simulations, we only aim to match the peak of model runs with the major peak of the observed energy deposition curve.

When a second energy deposition local peak occurs at a height above the peak energy deposition, the shock wave propagates downward and attenuates significantly, thus, does not have much of an effect on the ground $\Delta P$. However, if there is significant energy deposition at a height below the peak energy deposition height, this could be a source of uncertainty in determining the $\Delta P$. In that case, some of our estimates could be lower bounds.

The resulting predicted ground-level maximum $\Delta P$ is shown in Fig. 7. The model peak maximum $\Delta P$ is 740 Pa while the median $\Delta P$ is 500 Pa. These values bracket the observed light curve equivalent $\Delta P$ value of 590 Pa, but all are slightly lower than the Cart3D value of 937 Pa . The $\Delta P(500)$ is about 1300 km$^2$ as indicated by the dashed line in Fig. 7, very similar to the nominal Cart3D result. For comparison, at 500 Pa, typical sized windows (0.5-1.5 m$^2$) start to break at probability levels of $\sim$0.4 - 7%. Large size windows (2 m$^2$) would have a breakage probability of $\sim$12% at this pressure.

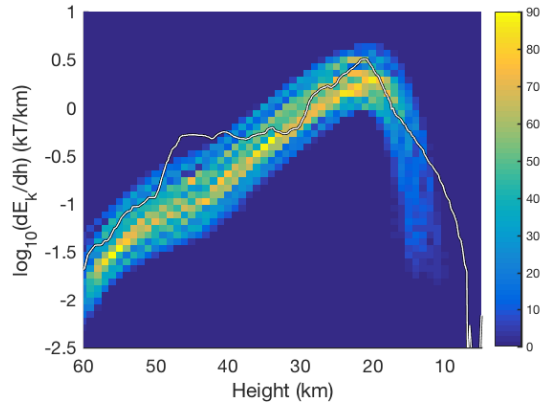

Figure 6: The TPFM model fits to the observed energy deposition curve for the Marshall Islands fireball. Out of 1000 simulated runs, we found 45 that match all of our empirical correlations and known observational constraints. All of the accepted model runs are plotted. The colorbar represents the number of energy deposition-height pairs that overlap in a particular pixel, where yellow shows highly populated number of runs clustered together. For the filtered simulated runs, the best-fit average initial mass was $4.2 \pm 1.5 \times 10^5$ kg, the average initial radius was $3.2 \pm 0.5$ m, and the average energy of 30 kT. This compares well with model results from Nemtchinov et al. (1997) who obtained a mass of $4 \times 10^5$ kg and an energy near 31 kT as well as a radius of 3.1 m using an analytic single-body pancake-type ablation model.

*Summary for five calibration fireballs*

The $\Delta P$ results from all five calibration events are summarized in Table 3. We compare the peak $\Delta P$ computed based on the observed light curves with median and maximum $\Delta P$ computed based on the modelled light curves. In all cases (except for the Antarctica event as discussed earlier) the light curve maximum $\Delta P$ lies between the median and maximum model ranges for $\Delta P$.

We also show the computed ground footprint in terms of the area (km$^2$) above which the expected median and maximum $\Delta P$ exceeded the 200 Pa and 500 Pa limits and compare with area footprints computed from the observed light curves. In general, our light curve derived $\Delta P$ values and ground area footprints found from TPFM models which are selected on the basis of the empirical criteria fits discussed earlier are within a factor of several as compared to the values which would be found using the actual light curve or from Cart3D

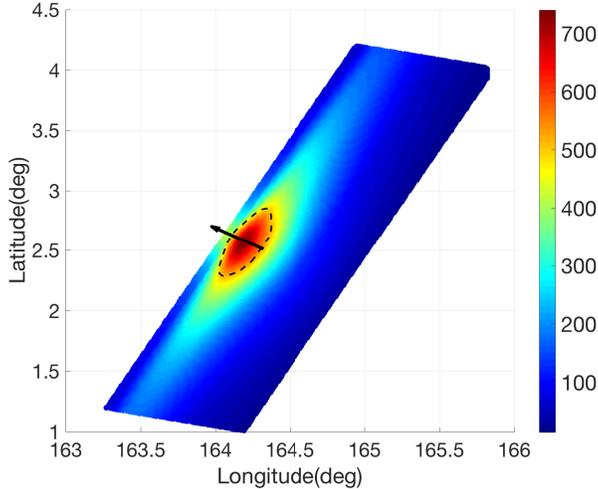

Figure 7: The Monte Carlo model predicted ground-level maximum weak shock $\Delta P$ (Pa) for the Marshall Islands fireball. The arrow represents the bolide trajectory from an altitude of 60 km to 10 km moving northwest. The colormap shows ground points reachable by the cylindrical shock during ablation between 60 and 10 km altitude at 1 km increments. The colorbar represents the $\Delta P$ and the dashed line shows the boundary inside of which the $\Delta P$ exceeded 500 Pa.

modelling, though in most cases these are all small areas.

On this basis, we believe that applying our generic Monte Carlo TPFM model approach constrained by empirical criteria to the entire suite of energetic JPL fireballs (all of which do not have available light curves) should yield reasonable limits on expected window breakage on the ground. We apply our formalism in the next section to this suite of bolides, examining both the expected peak maximum $\Delta P$, and ground $\Delta P$-Area footprints.

### JPL fireball events

Having validated our method by analyzing five fireballs where light curves are known, we next examined a number of energetic bolide events $(E > 2\,\text{kT})$ (SM section A Table S1) to estimate the characteristics of the resulting weak shock $\Delta P$ on the ground. As described earlier, for each event, we used the TPFM model to generate ranges of energy deposition with height, consistent with the speed, height of peak brightness, entry angle and energy reported on the JPL website. A total of 10,000 realizations were run for each event and a sub-set of the runs consistent with our empirical criteria were retained. The resulting energy deposition curves are then combined with the weak shock analytic model to compute ground $\Delta P$. In all, 18 fireballs reported on the JPL web page over the last 25 years had sufficient information and were above our threshold energy (2 kT) to allow modelling with our approach. The predicted maximum $\Delta P$ plots for each of the 18 events can be found in the SM section G.

| Event name | Value | Using observed light curves | Median $\Delta P$ of simulations | Maximum $\Delta P$ of simulations |
|---|---|---:|---:|---:|
| Marshall Islands | Peak $\Delta P$ | 590 | 500 | 740 |
| | $\Delta P(200)$ | 7200 | 5900 | 10000 |
| | $\Delta P(500)$ | 260 | 9 | 1300 |
| Tagish Lake | Peak $\Delta P$ | 230 | 150 | 240 |
| | $\Delta P(200)$ | 76 | - | 1200 |
| | $\Delta P(500)$ | - | - | - |
| Park Forest | Peak $\Delta P$ | 140 | 92 | 167 |
| | $\Delta P(200)$ | - | - | - |
| | $\Delta P(500)$ | - | - | - |
| Antarctic | Peak $\Delta P$ | 340 | 380 | 585 |
| | $\Delta P(200)$ | 3900 | 4800 | 13000 |
| | $\Delta P(500)$ | - | - | 510 |
| Tajikistan | Peak $\Delta P$ | 45 | 35 | 65 |
| | $\Delta P(200)$ | - | - | - |
| | $\Delta P(500)$ | - | - | - |

Table 3: Comparison of peak $\Delta P$ (Pa) and threshold $\Delta P$-areas (km$^2$) computed based on both the observed light curves and the simulated light curves for five calibration fireballs. Here $\Delta P(200)$ and $\Delta P(500)$ represent the ground-level areas where the $\Delta P$ exceeds 200 Pa and 500 Pa, respectively. For the simulation result, peak $\Delta P$ and threshold $\Delta P$-areas were computed based on the median $\Delta P$ plot (see SM section F) and the maximum $\Delta P$ plot (section Empirical ablation modelling: Five calibration case studies in the main text and SM section E), the latter providing an upper limit to the expected $\Delta P$.

Fig. 8 shows the peak median and maximum ground $\Delta P$ as a function of JPL fireball energy, color coded by (a) height (km) at the peak brightness and (b) entry angle (°). The height at the peak brightness makes the largest difference in peak $\Delta P$ for events of similar energy, as expected. Fireballs having as little as 5 kilotons of energy, if they penetrate to low enough heights ($< 26$ km), can produce $\Delta P$ on the ground in the half kilopascal range. For low energy events in general, we obtain higher $\Delta P$ with shallower entry angles also as expected. This is mainly because the minimum range to the ground for ballistic shocks is smaller than for steeper events, if all other quantities are the same.

We also calculated the total ground area (km$^2$) under the fireball trajectory where the maximum $\Delta P$ exceeded the 200 Pa and 500 Pa thresholds. Fig. 9 shows these $\Delta P$-area footprints color coded by the fireball (a) height (km) at the peak brightness and (b) entry angle (°). It is clear that more energetic events affect larger areas. All bolides having $E > 5$ kT produced peak maximum $\Delta P$ greater than 500 Pa. However, one event (2009-11-21) with very high height at peak brightness ($= 38$ km) produced peak maximum $\Delta P$ of only 390 Pa, even though it had a total energy of 18 kT. Similarly, all events with $E < 5$ kT produced lower than 500 Pa $\Delta P$, except one event (2003-09-27) that penetrated very deep into the atmosphere (height at the peak brightness $= 26$ km). This event had a maximum $\Delta P(500)$ of $\sim 10$ km$^2$.

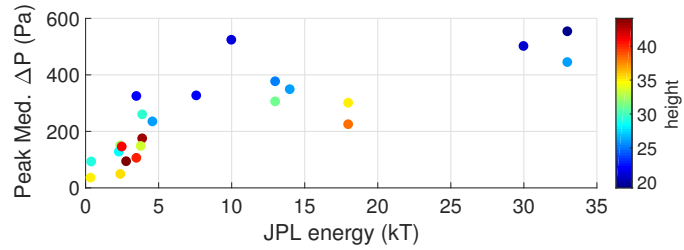
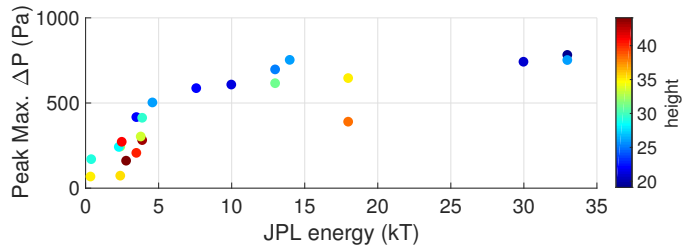

(a)

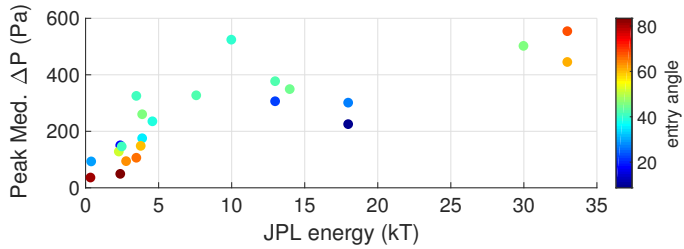
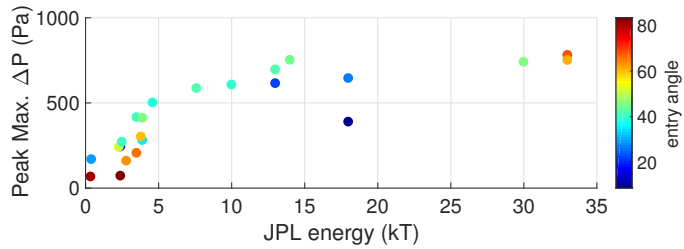

(b)

Figure 8: The predicted ground-level peak median and maximum $\Delta P$ for 18 of the most energetic JPL bolide events and 3 of our calibration fireballs having $E > 2\,\text{kT}$ as a function of energy. (a) Color represents the height (km) at peak brightness. (b) Color represents the entry angle with respect to the horizon.

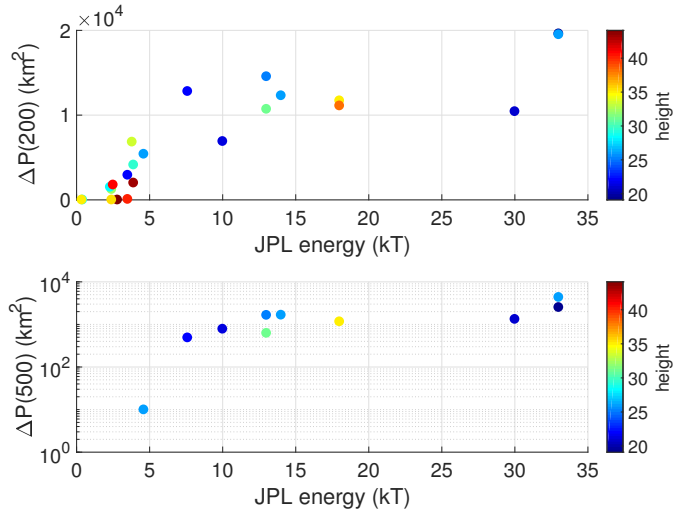

(a)

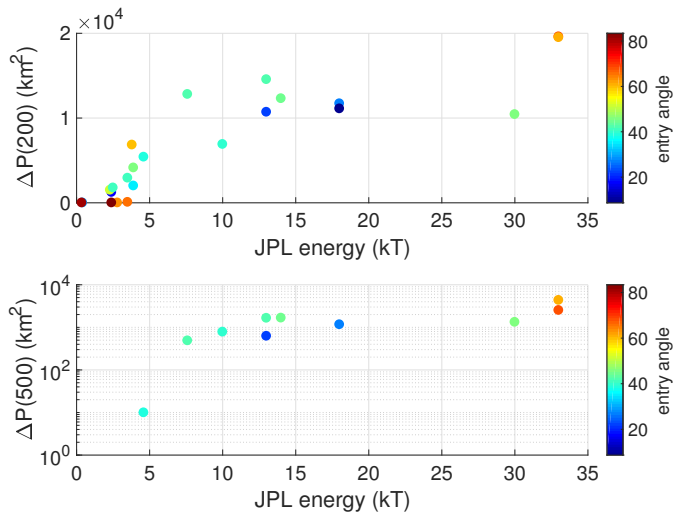

(b)

Figure 9: The calculated ground area (km$^2$) where the maximum $\Delta P$ exceeds 200 Pa and 500 Pa for 18 energetic bolide events and 3 calibration events (E > 2 kT) as a function of JPL energy. (a) Color represents the height (km) at the peak brightness. (b) Color represents the entry angle relative to the horizon.

**DISCUSSION**

In the following we focus on the maximum $\Delta P$ produced by our simulations. This represents the largest computed $\Delta P$ at each ground point across all realizations for a particular event and provides an upper limit from our simulations to the expected $\Delta P$.

Examination of Fig. 8 and 9 shows that the effective threshold energy at which fireballs in our case study produce $\Delta P$ levels where window damage would be heavy and might be reported (should these occur over an urban area) is $\sim$ 5 kT. That is, among the events we examined which occurred in the last quarter century globally, no fireball modeled with a JPL energy $<5$ kT had significant maximum ground $\Delta P$ in excess of 500 Pa using our simulation scheme. Virtually all fireballs having larger energy than this threshold produced maximum $\Delta P$ in excess of 500 Pa.

We note that in practice, our approach to modeling of a 5 kT JPL energy fireball encompasses an energy range up to 10 kT, as we have adopted a factor of two uncertainty in individual luminous efficient estimates following Nemtchinov et al. (1997). It is these highest energy realizations for a particular event which produce the maximum $\Delta P$ on the ground. Hence a more realistic limit on the total fireball energy required to produce window damage is $\approx 10$ kT. Based on the energy–impact frequency ranges for bolides given in Brown et al. (2002) and Brown et al. (2013), a 10 kT event impacts the Earth every 1-2 years.

The individual $\Delta P(200)$ and $\Delta P(500)$ for all 18 JPL events and our five calibration events can be found in the SM section G Table S4. As a reminder, our 200 Pa and 500 Pa were thresholds chosen for breaking a standard sized window (area of 0.5-1.5 m$^2$) with a probability of $\sim 0.01$-0.7% and 0.4-7% while a large window with area $>2$m$^2$ would have a breakage probability of $\sim 1.5$% and 12% at 200 Pa and 500 Pa, respectively.

From our examination of these 23 energetic fireballs occurring over the last 25 years we find the cumulative total surface area of the Earth that has experienced a maximum $\Delta P$ greater than 200 Pa from fireball shock waves was $1.6 \times 10^5$ km$^2$.

This translates approximately into an average annual affected area of $6.2 \times 10^3 \, \text{km}^2$. Similarly, the total affected ground area where maximum $\Delta P$ exceeded $500 \, \text{Pa}$ was $1.5 \times 10^4 \, \text{km}^2$, resulting in an average affected area of $580 \, \text{km}^2$ every year (Table 4).

However, our dataset of 23 fireballs (18 JPL events plus five calibration bolides) does not include the February 15, 2013 Chelyabinsk fireball. This is the largest recorded airburst on Earth since the 1908 Tunguska event (Brown et al., 2013; Popova et al., 2013). Unfortunately, we cannot apply our method to model this event at all ground points, as some ground points nearly under the fireball have geometry such that at one blast radius of range the atmospheric pressure changes by a factor of several, violating one of the assumptions in the use of the ReVelle weak shock model. The peak $\Delta P$ below the Chelyabinsk fireball, for example, is not determined with our approach, though the weak shock approach is marginally applicable to downtown Chelyabinsk as it has a comparatively large slant range from the peak energy deposition point on the trail.

Thus, we extracted the estimated $\Delta P(500)$ from Popova et al. (2013), where they used a numerical entry model to estimate the $\Delta P$ contours on the ground, and showed that $\Delta P(500)$ would be $\sim 1.9 \times 10^4 \, \text{km}^2$. This is more total ground area having $\Delta P(500)$ than all other fireballs in the last 25 years combined. This brings the average annual $\Delta P(500)$ affected area (including Chelyabinsk) to $\sim 10^3 \, \text{km}^2$. For comparison, the $\Delta P(1000)$ from Popova et al. (2013) was $\sim 10^4 \, \text{km}^2$ while Aftosmis et al. (2016), using the Cart3D model and a slightly different energy deposition profile to predict the ground overpressure footprint for the Chelyabinsk airburst found that $\Delta P(1000)$ was $\sim 2.0 \times 10^4 \, \text{km}^2$.

For the lower $\Delta P$ limit of $200 \, \text{Pa}$ at larger slant ranges (where most of the ground area is located) we can make use of the weak-shock model to produce an estimate for Chelyabinsk of $\Delta P(200)$, which we find to be $\sim 4.5 \times 10^4 \, \text{km}^2$. This brings the total annual $\Delta P(200)$ affected area (including Chelyabinsk) to $8 \times 10^3 \, \text{km}^2$.

For the more significant $\Delta P(500)$, the majority of the risk for window dam-age is caused by the very largest events, notably Chelyabinsk in our study time frame. This is as one would expect and is similar to the distribution of risk in the overall impact hazard, wherein the largest events cause the majority of the damage over the longest timescale (cf. Boslough et al. (2015)).

| Med. $\Delta P(200)$ | Med. $\Delta P(500)$ | Max. $\Delta P(200)$ | Max. $\Delta P(500)$ |
|---|---|---|---|
| $2.2 \times 10^3$ | 3.6 | $6.2 \times 10^3$ | 580 |

Table 4: Summary of annual total ground-level areas (km$^2$) where the median and maximum $\Delta P$ exceed the 200 Pa and 500 Pa thresholds.

The fraction of the Earth's total surface area covered by urban area is approximately 1% (Liu et al., 2014). Taking this as the effective area with significant numbers of windows we have roughly $5 \times 10^6$ km$^2$ of earth's surface covered by buildings/windows. We can calculate the expected annual probability that a fireball will occur over an urban area capable of producing ground-level $\Delta P$ at our 200 or 500 Pa threshold by the ratio of the urban area to the total surface area of the Earth compared to the maximum $\Delta P(200)$ or $\Delta P(500)$ areas.

Using current values for global urbanization, we expect an urban area to be affected once per $\approx 5000$ years by a fireball producing $\Delta P(500)$ where a single standard sized window would break at the 0.4 - 7% probability level. Similarly, roughly every 600 years we expect a fireball over an urban area producing a $\Delta P(200)$ with a probability of individual window breakage 0.01 - 0.7%. How many windows are actually broken for a given event depends on the details of the geometry of the fireball path relative to the urban area and peak $\Delta P$, but these values provide a guide to the expected intervals between major window-breaking fireball-produced events. Window breakage from fireballs should be a very rare occurrence. Viewed in this context, Chelyabinsk is an even more extraordinary event.

**CONCLUSIONS**

In this paper, we estimate how often fireballs produce window damage based on a case study of roughly two dozen energetic fireballs recorded in the last quarter century. This dataset consisted of 18 bolides (E > 2 kT) with limited flight data and 5 fireballs with light curves which we used to validate our entry model approach and the Chelyabinsk airburst. Our Monte Carlo entry modeling was used to estimate energy deposition curves which produced $\Delta P$ for the five validation events differing by no more than a few tens of percent from values computed using the actual light curves. In four of the five cases examined, the peak $\Delta P$ computed from the observed light curve fell between our model median and maximum peak $\Delta P$ computed using our generic Monte Carlo modeling approach. For one calibration case (the Sep 3, 2004 Antarctica fireball) we suggest the larger difference in observed vs. model $\Delta P$ results from the very low speed of the event and the correspondingly lower luminous efficiency (and hence higher total energy) for this event compared to the nominal JPL computed energy.

Based on the overall relationship between ground $\Delta P$ and bolide energy from all 23 of our simulated fireballs, we found that total impact energy plays the largest role in determining the ground-level $\Delta P$ with the height at the peak brightness and entry angle also affecting values. Given the same energy, the entry angle and the height at the peak brightness had the biggest effect on the ground overpressure. Bolides with lower height at the peak brightness produced higher $\Delta P$, affecting larger areas on the ground. Similarly, higher $\Delta P$ was typically obtained with shallower entry angle.

We find that fireballs with E $\sim$ 5 - 10 kT were needed to produce maximum $\Delta P$ greater than 500 Pa, which we would associate with heavy window damage on the ground in a dense urban area. At this $\Delta P$ level, window breakage occurs with a probability of 0.01 - 0.7% for standard sized windows (area of 0.5 - 1.5 m$^2$) and a probability of 0.4 - 7% for large windows (area > 2 m$^2$). This suggests that the effective threshold energy for fireballs to produce window damage is $\sim$5 - 10 kT, such events happening every 1-2 years globally.

Calculation of the equivalent average annual $\Delta P(500)$ and $\Delta P(200)$ based on all major fireball events (including Chelyabinsk) detected in the last 25 years produced annualized affected areas of $10^3$ and $8 \times 10^3 \, \text{km}^2$ respectively. This leads to an average recurrence interval for fireballs producing $\Delta P(500)$ over an urban area approximately once every 5000 years while the expected frequency of urban area exposure to fireballs producing $\Delta P(200)$ is every $\sim$600 years.

During our case study interval (1992-2017) a total of 18 fireballs were recorded with $E > 2 \, \text{kT}$ which had velocity, height and location information. Among these, the majority contribution to the total global areal $\Delta P$ footprint caused by their associated shocks producing $\Delta P(500)$ was from the Chelyabinsk fireball. The largest events dominate the long term damage at high $\Delta P$s. In contrast, Chelyabinsk was responsible for only about 1/4 of the cumulative areal ground exposure at the lower $\Delta P(200)$. Smaller more frequent events (and particularly more deeply penetrating fireballs) are significant contributors at these lower $\Delta P$s, near the threshold where sonic boom $\Delta P$s historically begin producing window damage reports in urban areas.

In summary, we expect window breakage from fireballs to be a very rare occurrence with likely intervals between urban areas exposed to significant fireball $\Delta P$, just capable of damaging windows, to be on the order of century timescales. The widespread window damage from Chelyabinsk is expected over an urban area on multi-millenium timescales, though this value is factor of several uncertain as the recurrence rate of Chelyabinsk class airbursts remains similarly uncertain. Nonetheless, our results further underscore the uniqueness of having even one Chelyabinsk airburst over an urban area in modern times.

Though these long average recurrence intervals are comforting, we also emphasize that our analysis suggests that the largest annually occurring bolides are capable of producing heavy window damage. Multi-kiloton bolide events (in the 5-10 kT range), should they occur over a major urban centre with large numbers of windows, can easily produce economically significant window damage.

**ACKNOWLEDGEMENTS**

The authors thank Natural Resources Canada, the Canada Research Chair program and the Natural Sciences and Engineering Research Council of Canada for funding support. This work was also supported by the Meteoroid Environment Office through NASA co-operative agreement NNX15AC94A. We are particularly grateful to Dr. Natalia Artemieva, Dr. Olga Popova and Dr. Donovan Mathias for very thoughtful and detailed reviews which greatly improved an earlier version of this manuscript.

Support for M. Aftosmis was provided through the NASA Planetary Defense Coordination Office (PDCO) in the Planetary Science Division of NASAs Science Mission Directorate. Computing resources were provided by the NASA High-End Computing Program through the NASA Advanced Supercomputing Division at Ames Research Center.

ReVelle D. O. 2005. Recent Advances in Bolide Entry Modeling: A Bolide Potpourri. *Earth, Moon, and Planets* 97:1–35. doi:`10.1007/s11038-005-2876-4`.

ReVelle D. O. 2007. NEO fireball diversity: Energetics-based entry modeling and analysis techniques. In Near Earth Objects, our Celestial Neighbors: Opportunity and Risk. edited by Valsecchi G. B., Vokrouhlický D., and Milani A. Cambridge: Cambridge University Press pp. 95–106.

ReVelle D. O., Edwards W., and Sandoval T. D. 2005. Genesis – An artificial, low velocity "meteor" fall and recovery: September 8, 2004. *Meteoritics & Planetary Science* 40(6):895–916. doi:`10.1111/j.1945-5100.2005.tb00162.x`.

Robertson D. K., and Mathias D. L. 2017. Effect of yield curves and porous crush on hydrocode simulations of asteroid airburst. *Journal of Geophysical Research: Planets* 122:599–613.

Rumpf C. M., Lewis H. G., and Atkinson P. M. 2017. Asteroid impact effects and their immediate hazards for human populations. *Geophysical Research Letters* 44:3433–3440.

Sakurai A. 1964. Blast Wave Theory (No. MRC-TSR-497). Technical Report, Wisconsin Univ-Madison Mathematics Research Center.

Seaman L. 1967. Response of windows to sonic booms. Technical Report, Stanford Research Institute, Menlo Park, California, USA.

Shuvalov V. V., Svetsov V. V., and Trubetskaya I. A. 2013. An estimate for the size of the area of damage on the Earth's surface after impacts of 10-300-m asteroids. *Solar System Research* 47:260–267. doi:`10.1134/S0038094613040217`.

Shuvalov V. V., and Trubetskaya I. A. 2007. Aerial bursts in the terrestrial atmosphere. *Solar System Research* 41:220–230. doi:`10.1134/S0038094607030057`.
45

Supplementary Material

## A. Basic fireball measurement inputs

Table S1 summarizes fireball measurement inputs of the 18 JPL events plus 3 calibration events (E > 2 kT) used in our simulation.

Table S1: Summary of the 18 energetic fireball events (E > 2 kT) and three of the five calibration events (highlighted in grey): the Marshall Islands fireball (February 1, 1994), the Tagish Lake fireball (January 18, 2000), the Antarctica fireball (September 3, 2004) which are given on the NASA JPL fireball website. Of all JPL fireball events that are > 2 kT, we selected only events that have data on velocity as well as height and geographic location at peak brightness. For the three calibration fireball events which had E > 2 kT, the height and velocity were taken from Tagliaferri et al. (1995), Brown et al. (2002b), and Klekociuk et al. (2005), respectively. The energy estimated on the JPL site follows the procedure described in Brown et al. (2002a). Entry angle is from the horizontal.

| Date (yyyy/mm/dd) / Time (UT) | Latitude (°N) | Longitude (°E) | Height (km) | Velocity (km/s) | Entry angle (°) | Energy (kT) |
|---|---|---|---|---|---|---|
| 1994-02-01 / 22:30 | 2.7 | 164.1 | 21 | 25 | 45 | 30 |
| 2000-01-18 / 16:43 | 60.3 | -134.6 | 32 | 15.8 | 17.8 | 2.4 |
| 2003-09-27 / 12:59 | 21 | 86.6 | 26 | 18.2 | 38.5 | 4.6 |
| 2004-06-05 / 20:34 | 1.3 | -174.4 | 43 | 19.5 | 34.5 | 3.9 |
| 2004-09-03 / 12:07 | -67.7 | 18.8 | 25 | 13 | 41.9 | 13 |
| 2004-10-07 / 13:14 | -27.3 | 71.5 | 35 | 19.2 | 27.2 | 18 |
| 2006-09-02 / 04:26 | -14 | 109.1 | 44.1 | 14.2 | 63.1 | 2.8 |
| 2009-02-07 / 19:51 | 56.6 | 69.8 | 40 | 15.4 | 65.7 | 3.5 |
| 2009-09-04 / 02:23 | 42.5 | 110 | 28.3 | 24 | 50.9 | 2.3 |
| 2009-10-08 / 02:57 | -4.2 | 120.6 | 19.1 | 19.2 | 67.5 | 33 |
| 2009-11-21 / 20:53 | -22 | 29.2 | 38 | 32.1 | 8.6 | 18 |
| 2010-07-06 / 23:54 | -34.1 | -174.5 | 26 | 15.7 | 43.9 | 14 |
| 2010-09-03 / 12:04 | -61 | 146.7 | 33.3 | 12.3 | 59.6 | 3.8 |
| 2010-12-25 / 23:24 | 38 | 158 | 26 | 18.1 | 60.9 | 33 |
| 2013-04-21 / 06:23 | -28.1 | -64.6 | 40.7 | 14.9 | 40.8 | 2.5 |
| 2013-04-30 / 08:40 | 35.5 | -30.7 | 21.2 | 12.1 | 39.5 | 10 |
| 2013-10-12 / 16:06 | -19.1 | -25 | 22.2 | 12.8 | 40.9 | 3.5 |
| 2014-05-08 / 19:42 | -36.9 | 87.3 | 35.4 | 19 | 83.4 | 2.4 |
| 2014-08-23 / 06:29 | -61.7 | 132.6 | 22.2 | 16.2 | 42.1 | 7.6 |
| 2015-09-07 / 01:41 | 14.5 | 98.9 | 29.3 | 21 | 45.4 | 3.9 |
| 2016-02-06 / 13:55 | -30.4 | -25.5 | 31 | 15.6 | 21.9 | 13 |

## B. TPFM input parameters and empirical constraints

This section describes the full range of initial parameters explored in the TPFM model for the 5 calibration events. As an example, we included TPFM runs that are filtered based on the empirical constraints (see section 3.2 in the main text) for the Marshall Islands fireball.

Table S2: Summary of the initial parameters used in the TPFM model for the five calibration fireball events. For the remaining 18 fireballs simulated in the study, the same generic parameters/ranges were used as starting points in the simulation (i.e. shape factor, amount of kinetic energy remaining at the end of the simulation, type of atmosphere, wake mode, porosity range and allowable number of fragments) while the event specific data (energy, velocity, height at peak brightness) are extracted from the JPL fireball table. The range of energy, porosity, strength, and number of fragments are generated randomly and are uniformly distributed within the given range.

|  | Marshall | Tagish Lake | Park Forest | Antarctica | Tajikistan |
|---|---|---|---|---|---|
| Total energy (kT) | 30 | 2.4 | 0.41 | 13 | 0.36 |
| Energy range (kT) | 15-60 | 1.2-4.8 | 0.25-1.0 | 6.5-26 | 0.18-0.72 |
| Velocity (km/s) | 24.5 | 15.8 | 19.5 | 13 | 14.3 |
| Entry angle (°E) | 45.4 | 17.8 | 61 | 41.9 | 80 |
| Height at peak brightness (km) | 21 | 32 | 29 | 25 | 35 |
| Porosity (%) | 0 - 95 | 40 - 95 | 0 - 95 | 0 - 95 | 0 - 95 |
| Strength (MPa) | 0.001-0.73 | 0.005-0.24 | 0.006-0.30 | 0.0005-0.12 | 0.006-0.15 |
| # of fragments | 1 - 1024 ||||| 
| Shape factor | 1.209 (Sphere) ||||| 
| Amount of $E_k$ remaining at end height | 1% ||||| 
| Atmosphere | Non-isothermal ||||| 
| Wake mode | Collective wake ||||| 

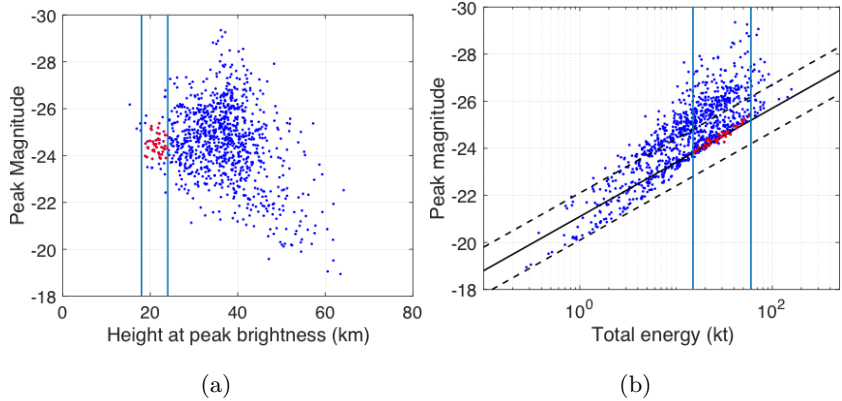

Figure S1: An example of a full set of TPFM runs that are filtered according to the observational constraints described in the main text, in this case for the Marshall Islands fireball. Blue circles are the 1000 simulated runs based on the initial parameters from Table B.6 while red circles are those runs filtered to match the observational constraints. The blue vertical solid lines show the range used for the observational constraints: (a) for the height at peak brightness of 18 - 24 km (nominal of $21 \pm 3$ km) and (b) the total energy range of 15 - 60 kT (nominal 30 kT). The black diagonal solid and dashed lines are the regression fit and the $2\sigma$ prediction intervals from the population as a whole for the correlation of peak brightness with total energy as shown in Fig. 3 in the main text.

Table S3: The range of parameters for the TPFM model runs which produce light curves that match our observational constraints and the empirical relations as described in the main text for the five calibration fireballs. The uncertainty in speed and entry angle was taken from the associated reference given in Table 2 in the main text. For the 18 JPL events ($E > 2$ kT), an uncertainty in speed of 0.5 km/s and uncertainty in entry angle of $1.0°$ were used. Compare to the full range of explored parameter space shown in Table S2.

|  | Marshall | Tagish Lake | Park Forest | Antarctica | Tajikistan |
|---|---|---|---|---|---|
| Velocity (km/s) | 21.5 - 26.7 | 14.6 - 16.9 | 18.7 - 20.1 | 12.8 - 13.3 | 12.8 - 15.8 |
| Entry angle (°E) | 43.9 - 47.7 | 14.6 - 21.9 | 26.4 - 31.7 | 40.0 - 44.0 | 76.7 - 82.7 |
| Porosity (%) | 2 - 55 | 40 - 75 | 1 - 67 | 1 - 71 | 1 - 83 |
| Strength (MPa) | 0.001 - 0.66 | 0.03 - 0.18 | 0.07 - 0.30 | 0.008 - 0.12 | 0.006 - 0.09 |
| # of fragments | 2 - 4 | 4 - 1024 | 2 - 1024 | 2 - 1024 | 4 - 1024 |

## C. Details of the Cart3D Modelling

The three dimensional modeling of blast propagation from airburst events with NASA's Cart3D simulation package is described in Aftosmis et al. (2016). The underlying solver is a second-order finite-volume method using a Cartesian cut-cell approach in which the governing equations for a compressible inviscid gas are discretized on a multilevel Cartesian mesh with embedded boundaries (Aftosmis and Berger, 1998). The solver has been extensively validated for aerodynamic flows and recent work presented a detailed description of this method applied to bolide entry simulations (Aftosmis et al., 2016). Computational meshes are comprised of multi-level Cartesian hexahedra which are clipped against the ground plane at the boundary.

The computational mesh setup was based upon experience with dozens of bolide entries and large blasts. The domain extended $250\,\text{km} \times 250\,\text{km} \times 80\,\text{km}$ (downrange × crossrange × altitude), and the meshes for the 1994 Marshall Islands and 2004 Antarctica fireballs used roughly 160 and 180 million elements (respectively). Mesh resolution near the entry corridor was $\sim 16\,\text{m}$, while along the ground plane resolution ranged from $16\,\text{m}$ near the peak ground overpressures to a coarser spacing of approximately $128\,\text{m}$ a distance $60\,\text{km}$ away.

The simulations were initiated with the time-dependent introduction of energy, momentum and mass following the deposition profiles shown in Fig. S2. The profile for the Feb 1, 1994 Marshall Islands event was based on Tagliaferri et al. (1995) while the Sept 3, 2004 Antarctica Fireball is from Klekociuk et al. (2005).

Fig. S3 presents an overview of the simulation through contour plots on the symmetry plane taken at two times during the evolution of the airblast. The upper frames show Mach contours while the lower frames show local overpressure. The snapshots are taken $38\,\text{s}$ and $79\,\text{s}$ after entry, and the peak energy deposition was at an altitude of $21\,\text{km}$ and downrange at $x = 4.3\,\text{km}$. The images at $38\,\text{s}$ (frames a & c) show that the shock surrounding the entry corridor is very nearly cylindrical in shape despite the peak in the energy deposition curve at 21 km altitude. This occurs since the increase in the rate of energy release approaching the peak is largely matched by the increasing ambient density and pressure as the bolide descends through the atmosphere. The images at 79 s (b & d) display snapshots of the shock system near peak ground overpressure ($x \approx -10$ km) where there is a prominent reflection of the incident shock. This reflection produces the highest overpressures seen in the ground footprint shown earlier on the left of Fig. 5.

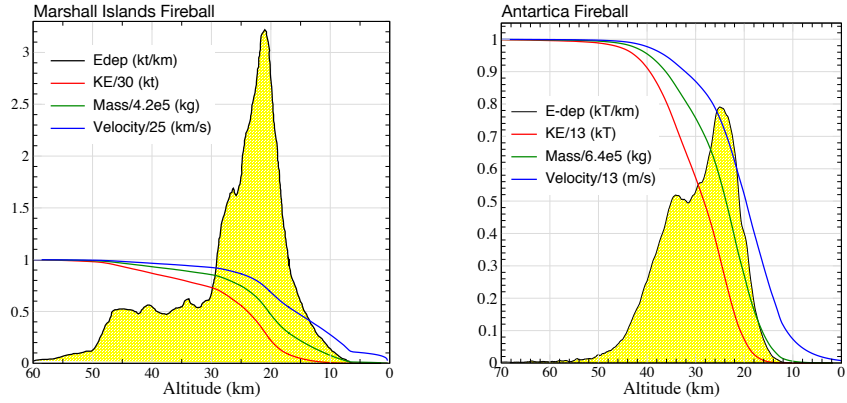

Figure S2: Energy deposition profiles used in Cart3D simulations of the Marshall Islands and Antarctica fireballs. Data from the Feb 1, 1994 Marshall Islands event was taken from Tagliaferri et al. (1995) while the Sept 3, 2004 Antarctica Fireball is from Klekociuk et al. (2005).

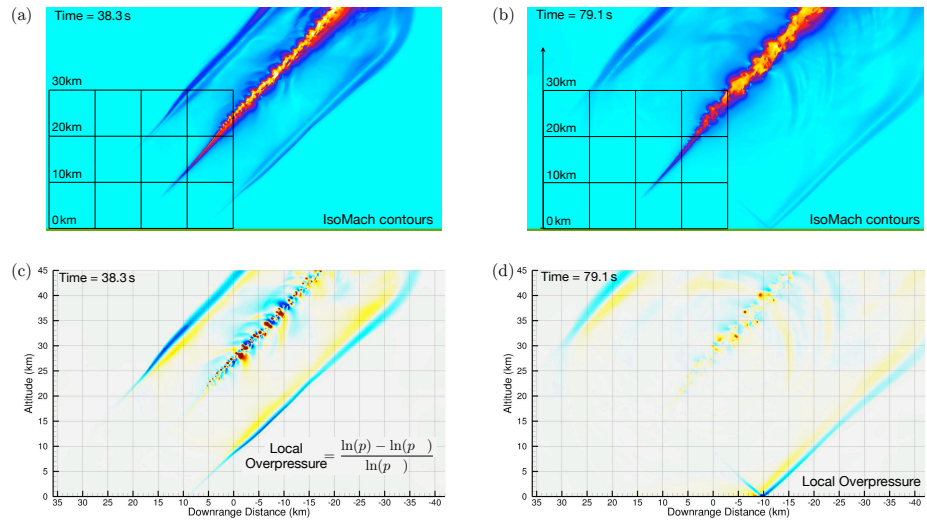

Figure S3: Overview of Cart3D simulations of 1994 Marshall Islands Fireball through symmetry plane Mach contours (a & b) and local overpressure (c & d). Frames on the left (a & c) show the cylindrical shock around the entry at 38.3 s after entry, while those at the right (b & d) show the intersection of the shock with the ground near the point of maximum ground overpressure around 79 s after entry.

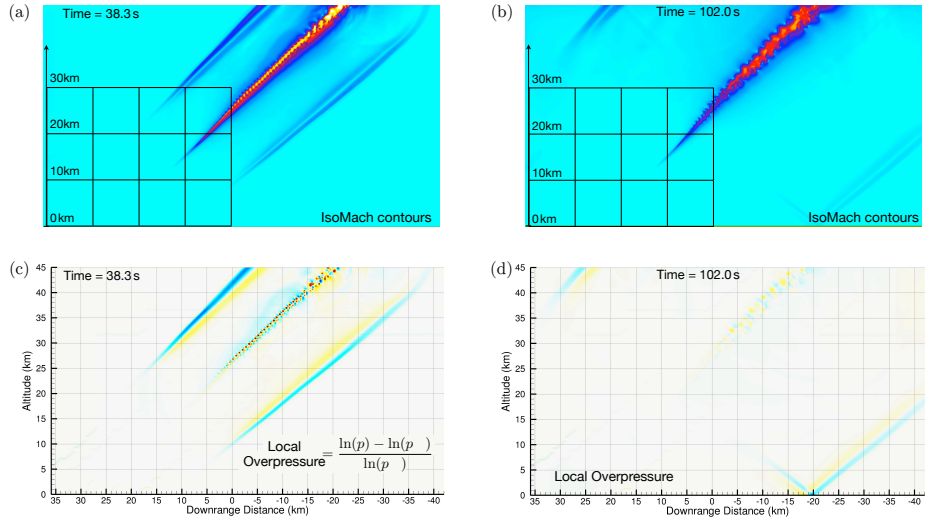

Figure S4: Overview of Cart3D simulations of 2004 Antarctica Fireball through symmetry plane Mach contours (a & b) and local overpressure (c & d). Frames on the left (a & c) show the cylindrical shock around the entry at 38.3 s after entry, while those at the right (b & d) show the intersection of the shock with the ground near the point of maximum ground overpressure around 102 s after entry.

Fig. S4 shows a similar set of snapshots of the Cart3D simulations of the 2004 Antarctica event. Owing chiefly to the low speed of this particular meteoroid, this event was only about one third as energetic as the Marshall Islands bolide. The peak energy deposition for this entry occurred at 25 km altitude and at $x = 0.0$ km downrange of the origin. The upper frames in Fig. S4 show Mach contours while the lower frames show local overpressure. The snapshots are taken 38 s and 102 s after entry. The images at 38 s (a & c) show that the shock system is very cylindrical with even less of a bulge near the peak in the energy deposition profile than the Marshall Islands simulation (Fig. S3(c)). This image also shows the beginnings of hydrostatic instability in the entry corridor itself as the flow in the hot wake becomes buoyancy driven. The snapshots at $102\,\text{s}$ (Figs. S4(b & d)) are taken near the time when the peak ground overpressure was recorded on the symmetry plane (near $x = -22\,\text{km}$, see right panels of Fig. 5). As before, this is a result of the strongest portion of the cylindrical shock reflecting off the ground plane.

### D. Conversion from the Light Curve to Energy Deposition Curve

For our five calibration fireball events, the raw data for the optical light curve was extracted from published figures in the references shown in Table 2 in the main text and converted to energy deposition using following process:

1. Convert the power digitized from the light curve to the energy per unit length:

$$E_l = \frac{4\pi P}{v} \qquad (8)$$

where $E_l$ is the optical energy per unit trail length (J/m), $P$ is the power (W/ster), and $v$ is the velocity (m/s).

2. Compute the total impact energy per unit length by dividing the optical energy per unit length by the energy efficiency, $\tau$ (Brown et al., 2002a):

$$\tau = (0.1212 \pm 0.0043) E_o^{0.115 \pm 0.075} \qquad (9)$$

where $E_o$ is the total optical radiant energy in kT ($1\,\text{kT} = 4.185 \times 10^{12}\,\text{J}$) provided from the JPL fireball dataset.

3. Calculate the total impact energy per unit height (J/m) by dividing the total impact energy per unit length (J/m) by the sine of the meteoroid entry angle.

The raw data of digitized light curve [time (sec) vs. power (W/ster)] and energy deposition curve [time(sec) vs. energy per unit height (kT/km)] for these five calibration events can be found here. More details concerning the instruments and analysis process can be found in Tagliaferri et al. (1994); Brown et al. (1995), Nemtchinov et al. (1997) and references therein.

### E. Empirical modelling: Four calibration case studies

We describe the four calibration events (the Tagish Lake, the Park Forest, the Antarctica, and the Tajikistan fireball) and compare the modelled energy deposition profiles to the observation. The predicted ground-level maximum weak shock $\Delta P$ plots for each event are also given.

*E.1. The January 18, 2000 Tagish Lake fireball*

This fireball occurred over northern Canada dropping C2 (ungrouped) meteorites on the frozen surface of Tagish Lake (Brown et al., 2000). The satellite optical light curve from Brown et al. (2002) was digitized and used to compare with our Monte Carlo TPFM model. The TPFM model fit to the observed energy deposition curve is shown in Fig. S5. The simulated runs do not reproduce the early light curve peak, but are a reasonable match to the lower altitude main light curve peak, which is our focus for ground level $\Delta P$ estimates.

We found that the average physical property values used in our model runs (as shown in Table 2) were very close to the initial physical properties of the meteoroid estimated in other studies. Hildebrand et al. (2006) bracketed the initial mass for Tagish Lake as between $6-9 \times 10^4$ kg based on short-lived radionuclide activities in recovered samples, while Brown et al. (2002) estimated a mass of $5.6 \times 10^4$ kg from entry modelling. ReVelle (2005) applied the TPFM and forward modelling to estimate an initial mass of $1.5 \times 10^5$ kg while Popova and Nemtchinov (2000) applied a single-body analytic pancake-type model to estimate an initial mass of 50-200 tonnes. These are all comparable to within a factor of two of our modelled mean initial mass of $9.5 \times 10^4$ kg. The predicted ground-level maximum $\Delta P$ is shown in Fig. S6 with the $\Delta P(200)$ of $1200\,\mathrm{km}^2$ bounded by a dotted line. The modeled peak maximum $\Delta P$ is 240 Pa, very close to the light curve-derived value of 230 Pa. This event is an order of magnitude less energetic than the Marshall Islands fireball and has a higher altitude maximum energy deposition. At 200 Pa, typical windows have a breakage probability between ∼0.01 - 0.7%. Though this event occurred over land, only a few struc-tures were within the 200 Pa contour, so the lack of reported window damage is unsurprising.

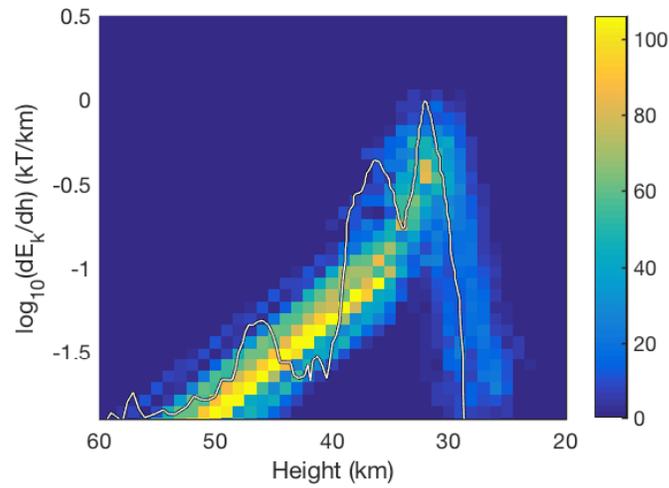

Figure S5: The TPFM model fit to the observed energy deposition curve for the Tagish Lake fireball. Out of 1000 simulated runs, we found 53 that match all of our empirical correlations and observational constraints. For the simulated runs, the average mass was $9.48\pm3.2\times10^4$ kg, the average radius was $2.4\pm0.3$ m, with an average energy of $2.8$ kT.

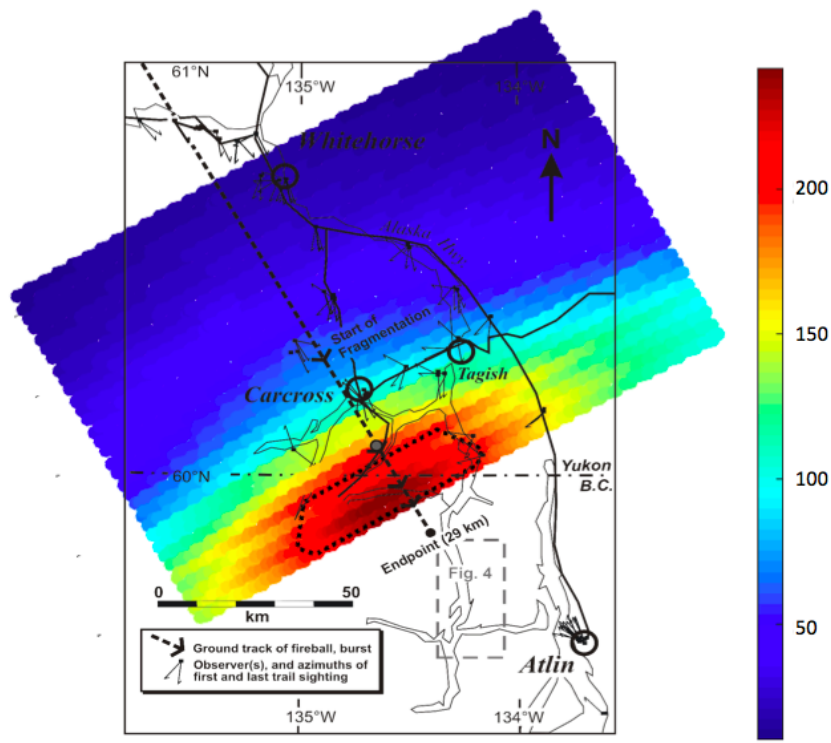

Figure S6: The predicted ground-level maximum weak shock $\Delta P$ (Pa) for the Tagish Lake fireball. The overlay map was taken from Hildebrand et al. (2006). In the map, the meteor moves southwestward, as shown with the dashed arrow line. The colormap shows all the ground points reacheable by the ballistic shock emanating from the trail between heights of 60 - 29 km. The dotted line shows the boundary where our predicted $\Delta P$ exceeds 200 Pa.

### E.2. The March 27, 2003 Park Forest fireball

This was the second lowest energy of all our calibration events, but of particular interest because it produced a large shower of L5 meteorites in an urban area (Simon et al., 2004). The Park Forest meteorite fall is likely the largest meteorite shower to occur in a modern urban setting.

Fig. S7 shows the TPFM simulated runs compared with the observed energy deposition curve. The observed curve shows three distinct peaks caused by fragmentation events at heights of 37, 29, and 22 km. The two fragmentation events at 37 and 22 km are not reproduced with our model runs; however the ensemble of simulations generally reproduce the observed maximum energy deposition (within a factor of two) at $\sim$29 km. Similarly, the average peak magnitude of the modelled fireball was -21.4, a good match to the observed peak absolute visual magnitude of -22 (Brown et al., 2004). Our mean modelled initial mass of $1.04 \times 10^4$ kg is similar to the estimate from Brown et al. (2004) but a factor of 2-3 higher than a more recent estimate by Meier et al. (2017) based on short-lived radionuclide or the estimate from ReVelle (2005). The model result (Fig. S8) suggests that the peak maximum $\Delta P$ was only 167 Pa, below the limit where reports of window damage even in a dense urban area, might be expected (e.g. Clarkson and Mayes (1972)). This also compares favorably with a peak maximum $\Delta P$ of 140 Pa computed from the actual light curve.

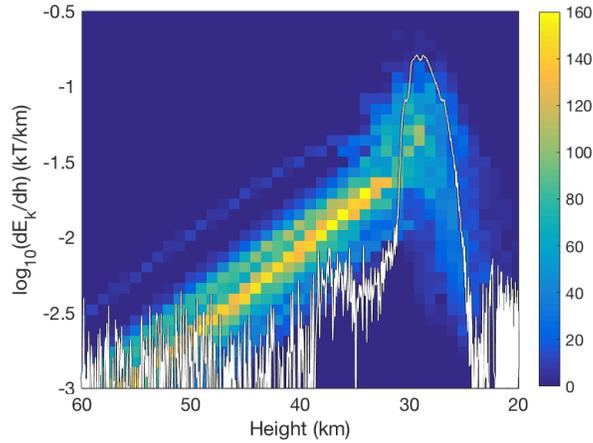

Figure S7: The TPFM model fit to the observed energy deposition curve for the Park forest fireball. The 80 runs which met all empirical criteria (as described in the text) are shown as color curves. For the simulated runs, the average mass was $10.4 \pm 4.0 \times 10^3$ kg, the average radius was $0.97 \pm 0.2$ m, with the average energy was 0.47 kT.

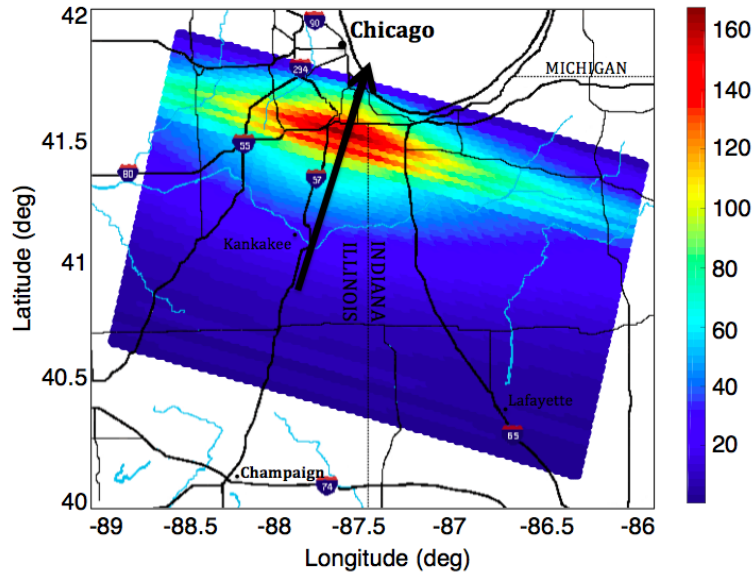

Figure S8: The predicted ground-level maximum weak shock $\Delta P$ (Pa) for the Park Forest fireball. The arrow represents the bolide trajectory from a height of 80 km to 18 km moving north-northeast. The colormap shows all the ground points that were accessible to the ballistic shock wave in this height interval. With the peak maximum $\Delta P$ of $\sim$167 Pa, there is less than a 0.1% probability of breaking typical windows in urban areas.

### E.3. The September 3, 2004 Antarctica fireball

The optical light curve for this fireball was measured by Department of Energy space-based visible light sensors and showed two major fragmentation episodes at altitudes of 32 km and 25 km (Klekociuk et al., 2005). The observed light curve, converted to an equivalent energy deposition curve (see SM section D for details) and compared with the TPFM model fit is shown in Fig. S9. The majority of our simulated runs produced about 4 times larger peak energy deposition than that derived directly from the observed light curve. This might be interpreted as the meteoroid being stronger or undergoing less fragmentation than a typical meteoroid; it may also be related to its very low entry speed. We emphasize that TPFM treats macroscopic fragmentation only; for events with significant production of small fragments/dust (e.g. Borovička et al. (2017)) the shape of the energy deposition profile will be modified. More complex models such as the Fragment Cloud Model of (Wheeler et al., 2017) can account for dust and we expect will provide better light curve fits in individual cases. However, for our fireball dataset which lacks light curve information we are using the simplest model that can match our population-wide empirical constraints to produce equivalent energy deposition curves near maximum.

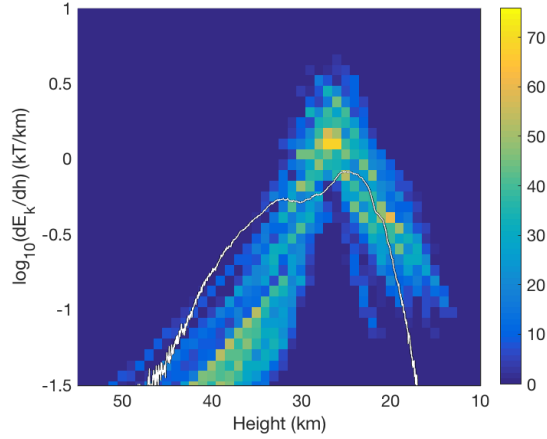

Figure S9: The TPFM model fit to the observed energy deposition curve (white line) for the Antarctica fireball. Out of 1000 simulated runs, we found 38 runs that match all our empirical correlations and observational constrains. For the simulated runs, the average mass was $6.04 \pm 2.0 \times 10^5$ kg, the average radius was $3.9 \pm 0.6$ m, with the average energy of 12.2 kT. Our model result matches well with Klekociuk et al. (2005) who obtained a total initial energy of 13 kT corresponding to a mass of $6.5 \pm 0.5 \times 10^6$ kg by applying entry modelling of the light curve and trajectory data.

The ground footprint associated with the maximum weak shock model is shown in Fig. S10 with the $\Delta P(500)$ inside the dashed line. The modeled peak maximum $\Delta P$ is 585 Pa, noticeably higher than the $\Delta P$ of 340 Pa found using the actual light curve . The model $\Delta P(500)$ corresponding to the area that would have experienced window breakage is about $510\,\mathrm{km}^2$. This fireball shows the greatest deviation between $\Delta P$ levels computed from the true light curve and energy depositions produced from our Monte Carlo modeling approach. It is also the lowest speed of all five of our calibration events. This emphasizes the potential limitations of our approach when applied to unusual or rare fireball populations, such as low speed events, which are not necessarily well represented in the population as a whole from which our empirical constraints are drawn.

We suggest this bias may reflect the fact that the true luminous efficiency is much lower at low speeds (Nemtchinov et al., 1997) than is assumed in the nominal JPL energy estimates when using the actual light curve. This is because the Brown et al. (2002) formulation for luminous efficiency used to compute JPL energies does not explicitly account for changes in luminous efficiency at low speeds but uses population averages. Hence the $\Delta P$ computed from the light curve for such low speeds would actually be too small, as we see for the Antarctica event.

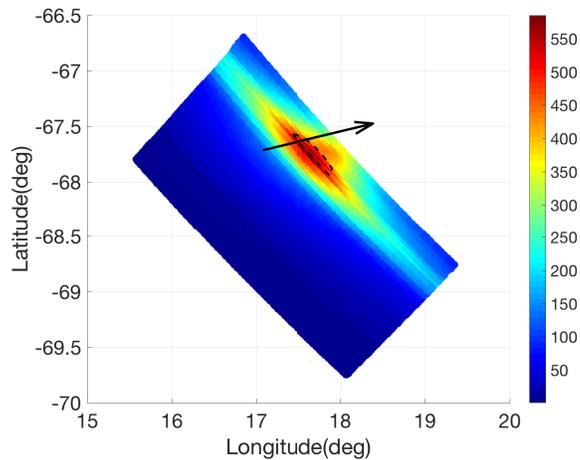

Figure S10: The predicted ground-level maximum weak shock $\Delta P$ (Pa) for the Antarctica fireball. The arrow represents the bolide trajectory from an altitude of $70\,\mathrm{km}$ to $16\,\mathrm{km}$ moving towards the east-northeast. The colormap shows all the ground points that were accessible by the cylindrical shock produced during ablation. The dashed line shows the boundary where the model maximum $\Delta P$ exceeds $500\,\mathrm{Pa}$.

### E.4. The July 23, 2008 Tajikistan fireball

The satellite optical light curve from Konovalova et al. (2013) was digitized and used to produce an equivalent energy deposition curve. The TPFM model runs are compared with the observed edep curve in Fig. S11. Our simulated runs only match the second and third flares of the observed energy deposition curve at 26 and 24 km to within a factor of 2 - 3. The maximum energy deposition from model runs and observations were in good agreement at a height of 35 km.

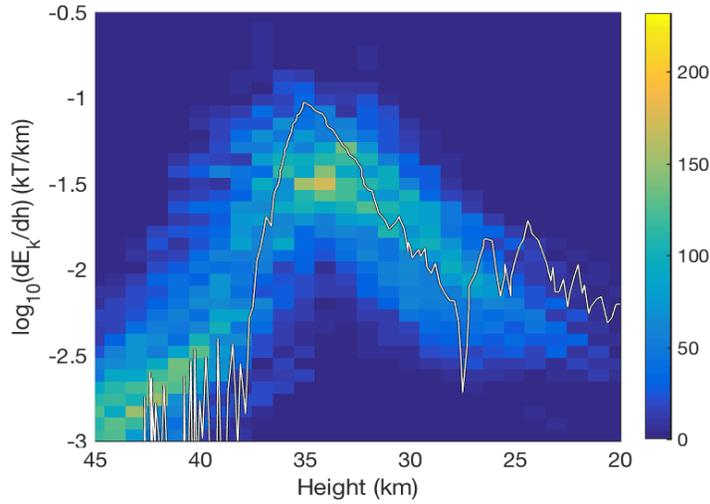

Figure S11: The TPFM model energy deposition profiles compared to the observed energy deposition curve for the Tajikistan superbolide. Out of 1000 simulated runs, we found 116 runs that match all of our empirical correlations and observational constrains. For the simulated runs, the average mass was found to be $1.4 \pm 0.4 \times 10^4$ kg, the average radius $1.1 \pm 0.2$ m while the average energy was 0.34 kT. Our model result shows a reasonable agreement with Konovalova et al. (2013) where they computed an initial mass of $20 - 25$ tons based on the theoretical estimates of initial kinetic energy of 0.59 kT.

Fig. S12 shows the result from the weak shock model indicating a peak maximum $\Delta P$ of 65 Pa and a median of 35 Pa, similar to the 45 Pa computed from the light curve and in all instances clearly well below levels that could produce window damage.

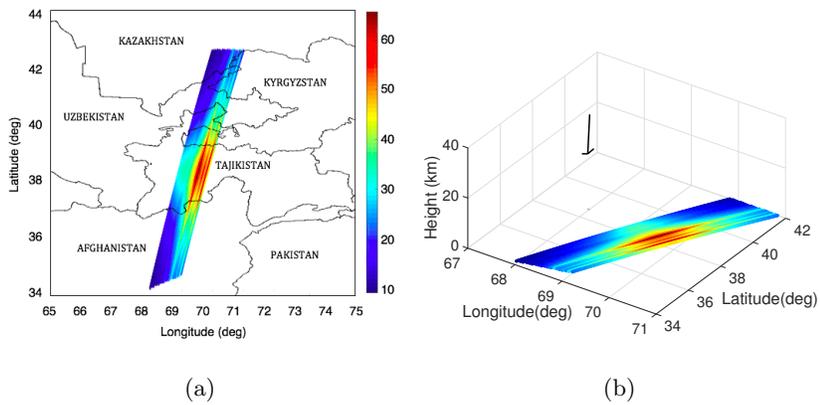

Figure S12: The predicted maximum ground-level weak shock $\Delta P$ (Pa) for the Tajikistan superbolide. (a) Top down view. The short black horizontal line at 38.5°N, 68°E indicates the bolide trajectory moving towards west. As the fireball entered the atmosphere at a very steep angle (80° from the horizontal), the ground projected length of the trajectory was very short, only ∼5.3 km. (b) 3D view. The arrow represents the bolide trajectory from a height of 38 km to 20 km.

## F. Details of the TPFM and the Weak Shock Model Results for Calibration Events

We present detailed plots showing our filtered model runs for all five calibration fireball events. These detailed model solutions are our primary means of validating our generic approach to estimating the energy deposition as a function of height for our complete suite of fireballs (where light curves are generally not available).

### F.1. The Marshall Islands Fireball

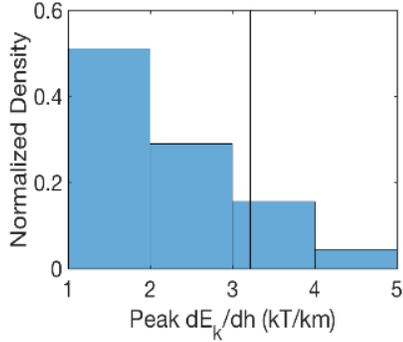

(a)

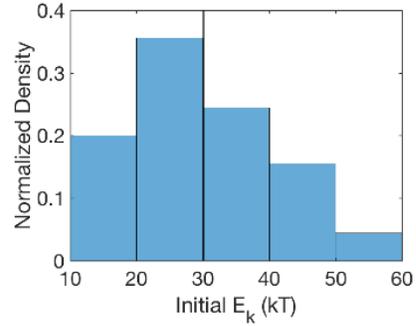

(b)

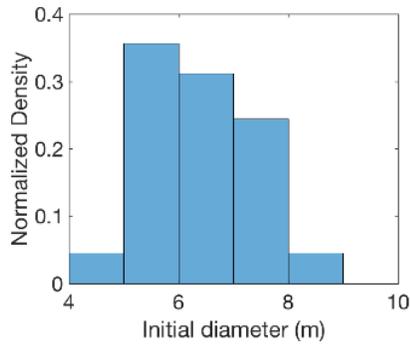

(c)

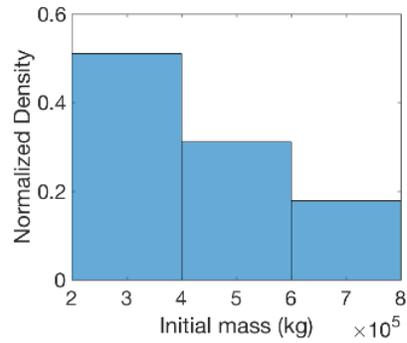

(d)

Figure S13: (a-d) Histograms showing the distribution of filtered TPFM model runs for 4 different parameters: (a) peak energy deposition per unit height, (b) initial kinetic energy, (c) initial diameter, and (d) initial mass. The vertical solid line corresponds to the observed quantity. The average of the peak energy deposition for the simulated runs is 2.1 kT/km and the observed peak energy deposition is 3.1 kT/km. The average of the initial kinetic energy for the ensemble of filtered model runs of (29.98 kT) was in a good agreement with the estimated JPL initial kinetic energy, (30 kT).

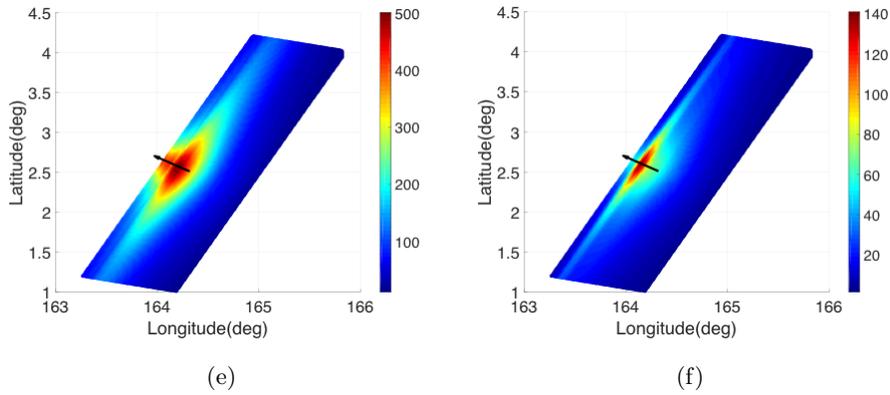

Figure S13: (e-f) The resulting overpressure predicted by the weak shock model based on the energy deposition curves produced from the filtered TPFM models. Shown are the median (e) and standard deviation (f) of the weak shock overpressure (Pa) for the Marshall Islands fireball. The arrow represents the bolide trajectory from 60 to 10 km altitude moving northwest.

*F.2. The Tagish Lake Fireball*

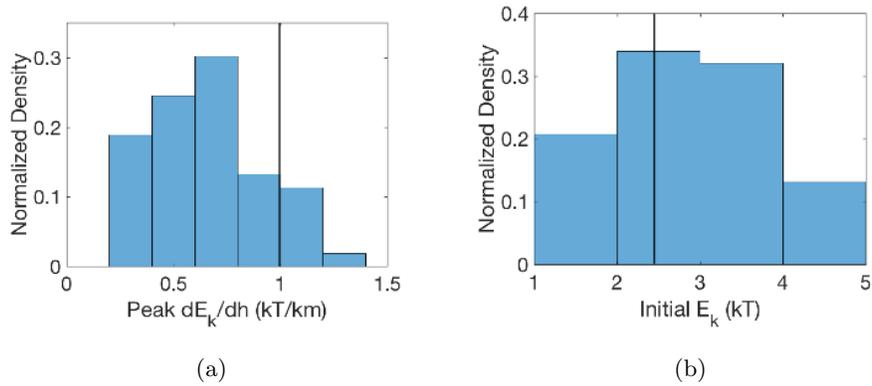

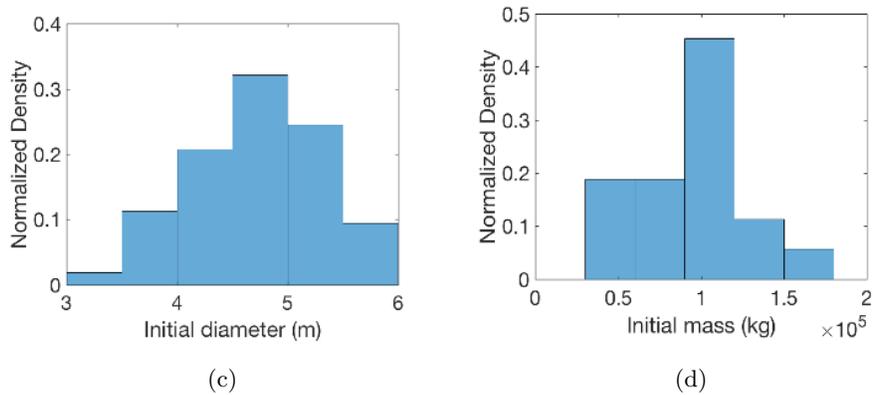

Figure S14: (a-d) Histograms showing the distribution of TPFM model runs for 4 different parameters: (a) peak energy deposition per unit height, (b) initial kinetic energy, (c) initial diameter, and (d) initial mass. The vertical solid line corresponds to the observed quantity. The average of the peak energy deposition per unit height for the simulated runs is 0.7 kT/km and the observed peak energy deposition is 1 kT/km. The average of model runs of initial kinetic energy is 2.8 kT and the estimated JPL initial kinetic energy is 2.4 kT.

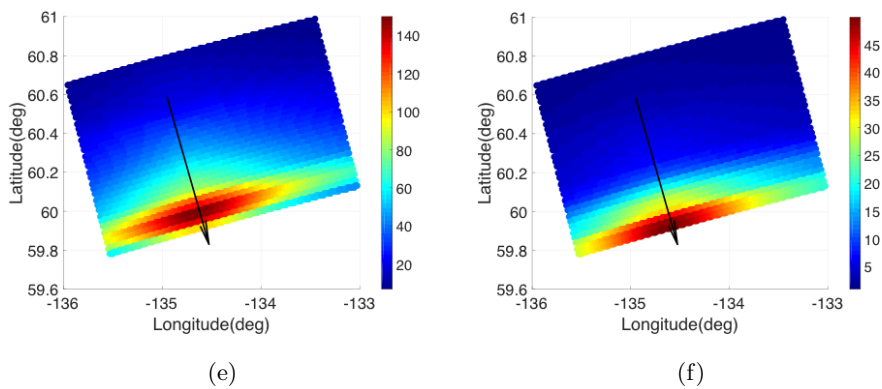

Figure S14: (e-f) The result of weak shock modeling showing the median (e) and standard deviation (f) of weak shock overpressure (Pa) for the Tagish Lake fireball. The arrow represents the bolide trajectory from 60 to 29 km moving southeast.

*F.3. The Park Forest Fireball*

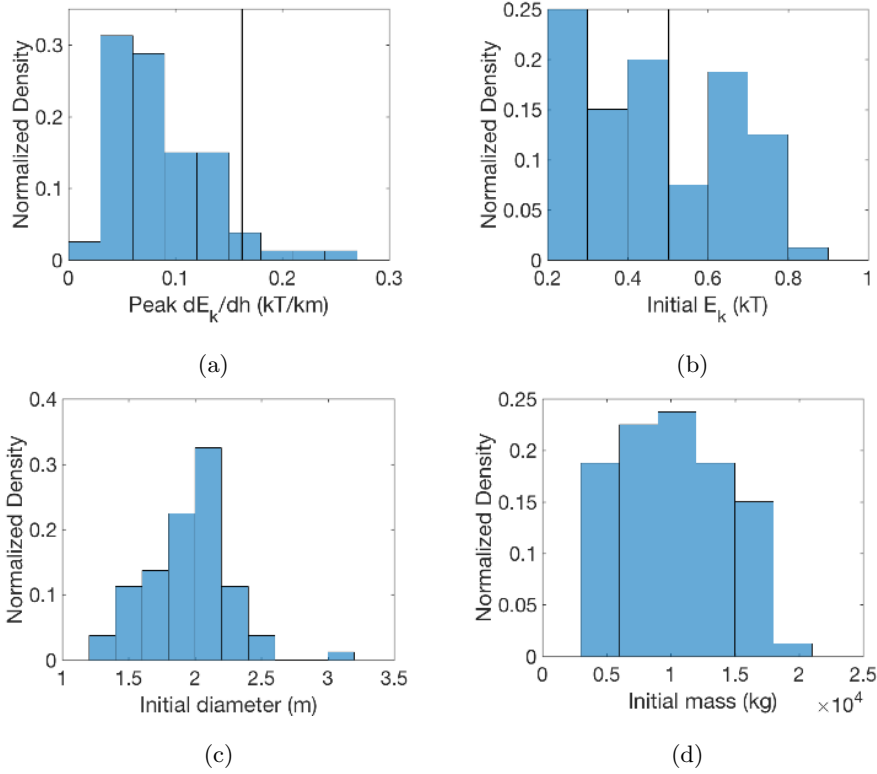

Figure S15: (a-d) Histograms showing the distribution of TPFM model runs for 4 different parameters: (a) peak energy deposition per unit height, (b) initial kinetic energy, (c) initial diameter, and (d) initial mass. The vertical solid line corresponds to the observed quantity. The average of the peak energy deposition per unit height for the simulated runs is $0.087\,\text{kT/km}$ and the observed peak energy deposition is $0.17\,\text{kT/km}$. The average of model runs of initial kinetic energy is $0.47\,\text{kT}$ and the estimated JPL kinetic energy is $0.41\,\text{kT}$.

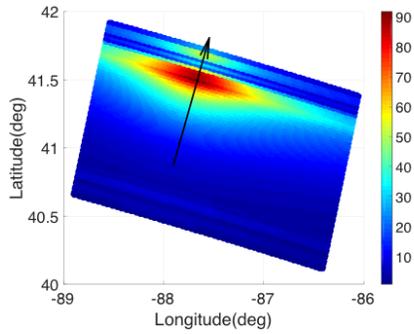
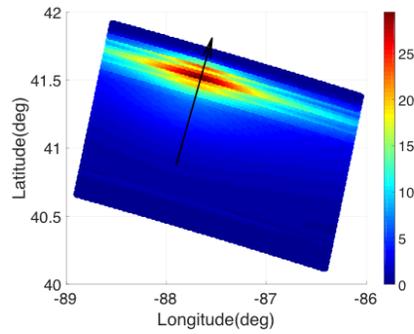

(e)                                 (f)

Figure S15: (e-f) The result of weak shock model showing the median (e) and standard deviation (f) of weak shock overpressure (Pa) for the Park Forest fireball. The arrow represents the bolide trajectory from an altitude of 80 to 18 km moving north-northeast.

*F.4. The Antarctica Fireball*

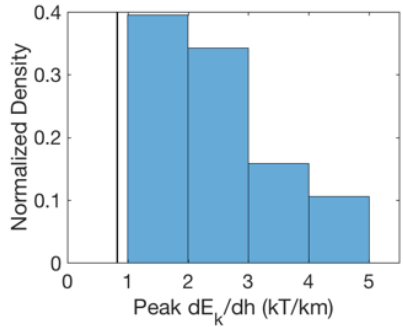
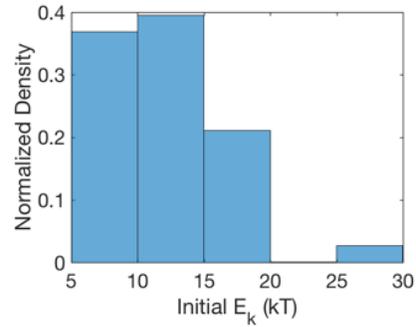

(a)                                 (b)

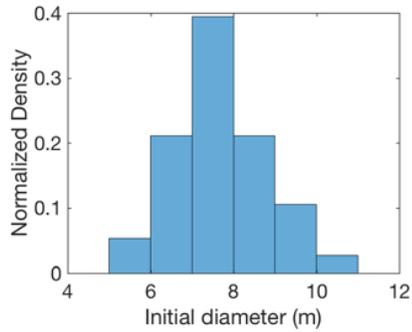
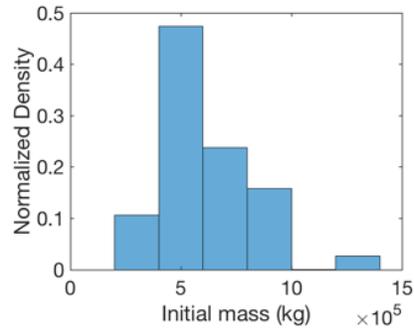

(c)                                 (d)

Figure S17: (a-d)Histograms showing the distribution of TPFM model runs for 4 different parameters: (a) peak energy deposition per unit height, (b) initial kinetic energy, (c) initial diameter, and (d) initial mass. The vertical solid line corresponds to the observed quantity. The average of the peak energy deposition per unit height for the simulated runs (2.5 kT/km) was not in good agreement with the observed peak energy deposition (0.9 kT/km), as the majority of our simulated runs showed about 4 times larger peak energy deposition than the observation. The average of model runs of initial kinetic energy is 12.2 kT and the estimated JPL kinetic energy is 13 kT.

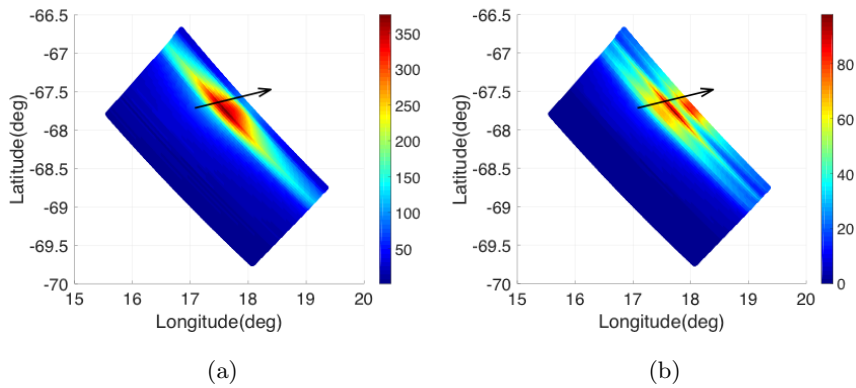

Figure S17: (e-f) The result of weak shock model showing the median (f) and standard deviation (g) of weak shock overpressure (Pa) for the Antarctica fireball. The arrow represents the bolide trajectory from 70 to 16 km moving east-northeast.

*F.5. The Tajikistan Superbolide*

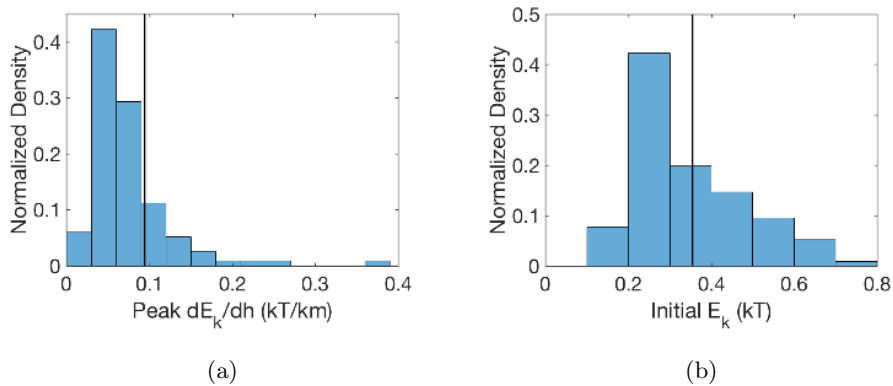

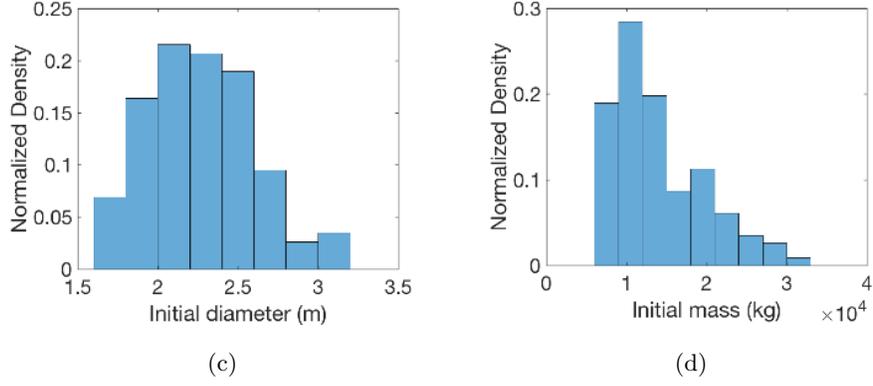

Figure S18: (a-d)Histograms showing the distribution of TPFM model runs for 4 different parameters: (a) peak energy deposition per unit height, (b) initial kinetic energy, (c) initial diameter, and (d) initial mass. The vertical solid line corresponds to the observed quantity. The average of the peak energy deposition per unit height for the simulated runs is 0.075 kT/km and the observed peak energy deposition is 0.09 kT/km. The average of model runs of initial kinetic energy is 0.34 kT and the estimated JPL initial kinetic energy is 0.36 kT.

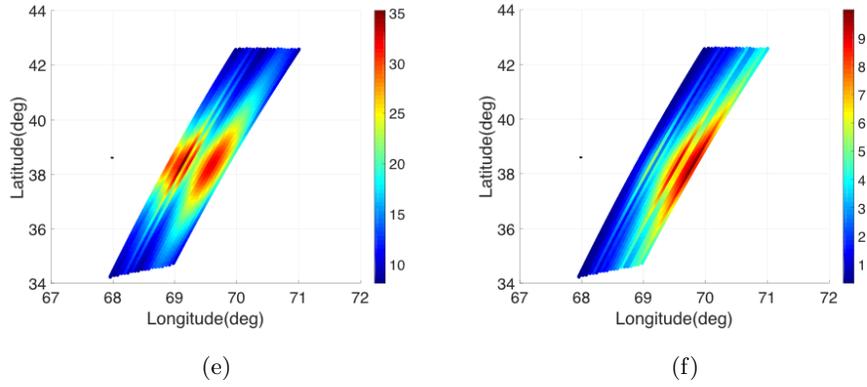

Figure S18: (e-f) The result of weak shock model showing the median (e) and standard deviation (f) of weak shock overpressure (Pa) for the Tajikistan superbolide. A short black horizontal line at 38.3N, 68E indicates the bolide trajectory from 38 to 20 km moving west.

## G. Weak Shock Model Results for 18 JPL Fireball Events

The predicted maximum $\Delta P$ plots for each of the 18 events are shown in this section. A summary of the areal footprint on the ground where $\Delta P$s exceeded $200\,\text{Pa}$ and $500\,\text{Pa}$ ($\Delta P(200)$ and $\Delta P(500)$) for all 18 JPL events and our five calibration events can be found in Table S4.

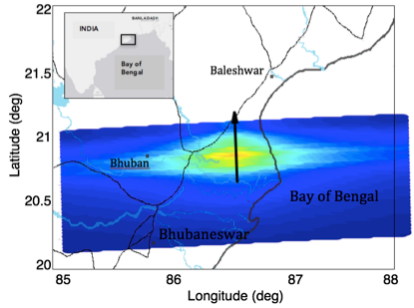
(a) 2003-09-27, India (4.6kT)

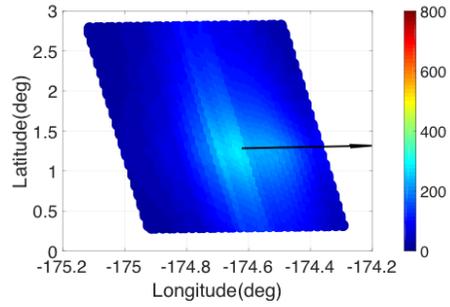
(b) 2004-06-05, N. Pacific Ocean (3.9kT)

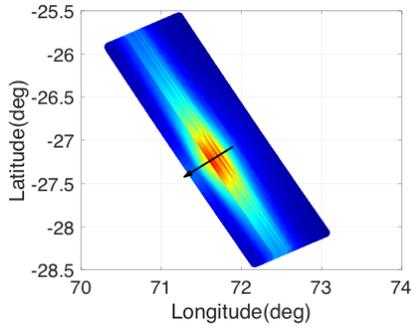
(c) 2004-10-07, Indian Ocean (18kT)

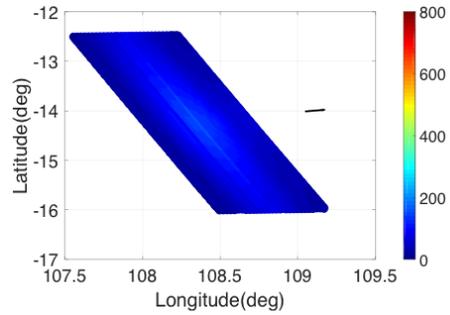
(d) 2006-09-02, Indian Ocean (2.8kT)

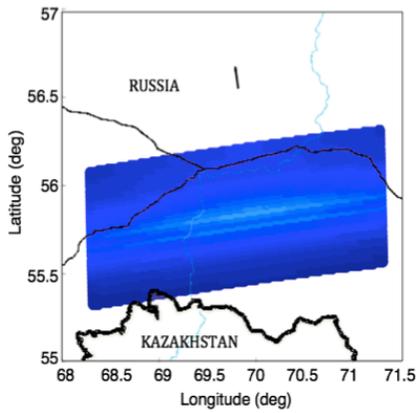
(e) 2009-02-07, Russia (3.5kT)

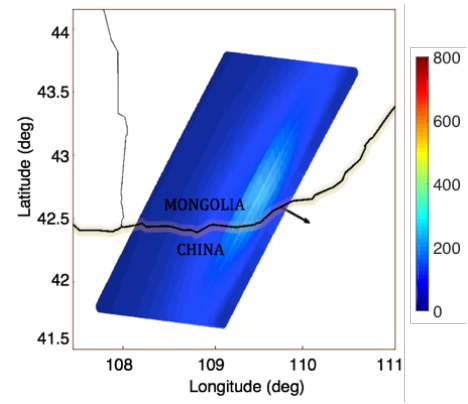
(f) 2009-09-04, China (2.3kT)

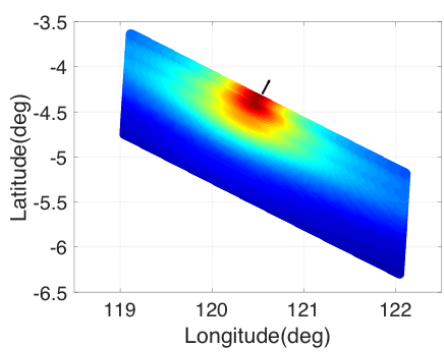
(g) 2009-10-08, Banda Sea (33kT)

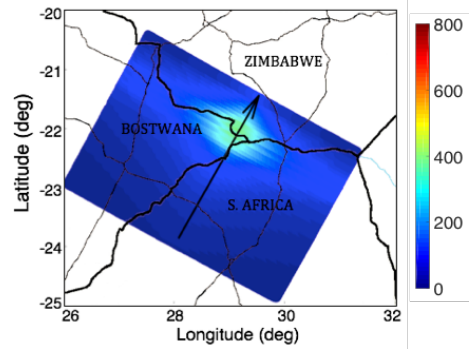
(h) 2009-11-21, Zimbabwe (18kT)

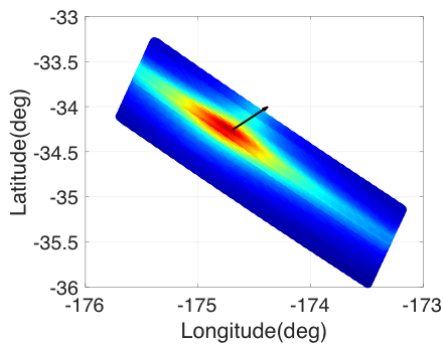
(i) 2010-07-06, S. Pacific Ocean (14kT)

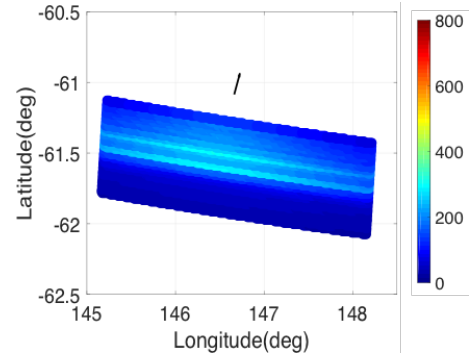
(j) 2010-09-03, Southern Ocean (3.8kT)

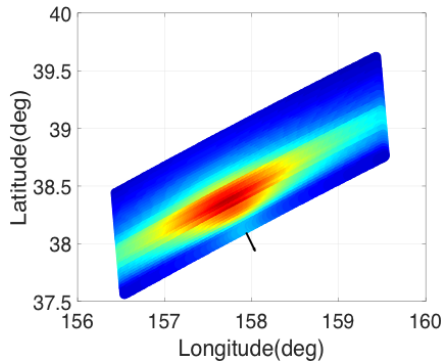
(k) 2010-12-25, N. Pacific Ocean (33kT)

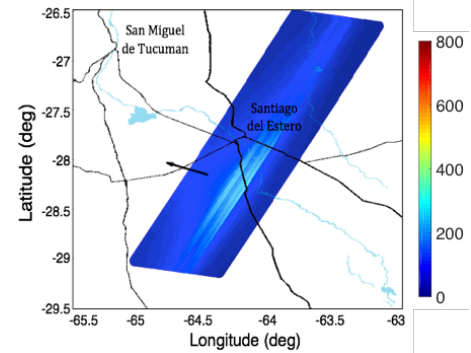
(l) 2013-04-21, Argentina (2.5kT)

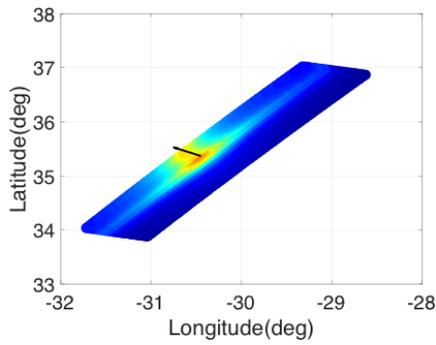 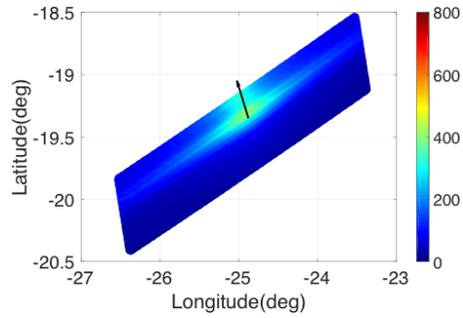

(m) 2013-04-30 , N. Pacific Ocean (10kT)   (n) 2013-10-12, S. Atlantic Ocean (3.5kT)

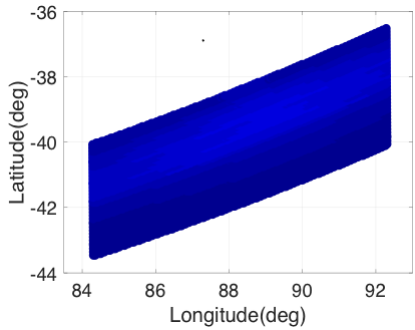 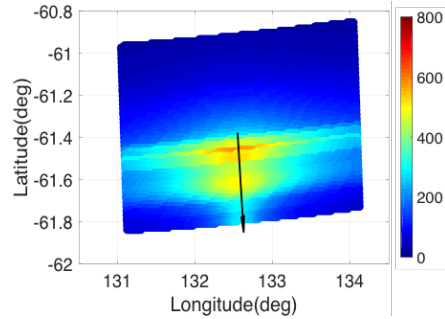

(o) 2014-05-08, Indian Ocean (2.4kT)   (p) 2014-08-23, Southern Ocean (7.6kT)

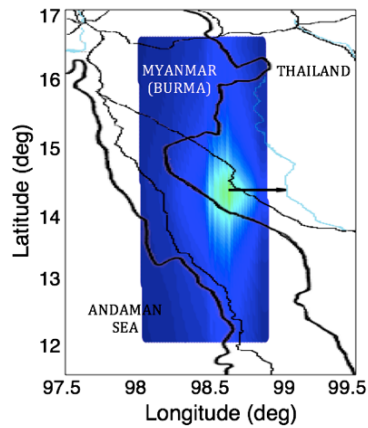 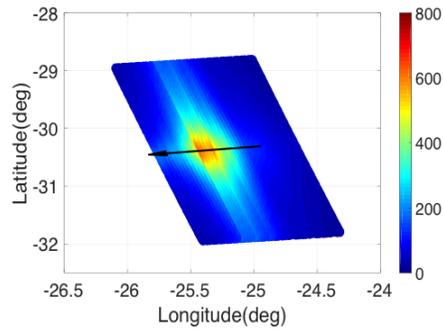

(q) 2015-09-07, Thailand (3.9kT))   (r) 2016-02-06, S. Atlantic Ocean (13kT)

Figure S19: The result of weak shock model showing the maximum overpressure (Pa) for 18 JPL fireball events. The arrow represents the bolide trajectory. The map was overlaid for the events that occurred over the land. Country border line (thick black line) with major roads/highways (thin black line) and major rivers (blue line) are shown.

Table S4: Summary of ground-level areas ($10^3\,\text{km}^2$) under the fireball where the median and maximum $\Delta P$ exceeded the 200 Pa and 500 Pa thresholds for 18 JPL fireball events and 5 calibration events (highlighted in grey). Max. $\Delta P(200)$ of the February 15, 2013 Chelyabinsk fireball (second last row) was computed following our modelling approach based on the energy deposition profile given by Brown et al. (2013), while Max. $\Delta P(500)$ was extracted from Popova et al. (2013).

| Date | Height (km) | Energy (kT) | Med. $\Delta P(200)$ | Med. $\Delta P(500)$ | Max $\Delta P(200)$ | Max $\Delta P(500)$ |
|---|---|---|---|---|---|---|
| 1994-02-01 | 21 | 30 | 5.9 | 0.01 | 10 | 1.3 |
| 2000-01-08 | 32 | 2.4 | - | - | 1.2 | - |
| 2003-03-27 | 29 | 0.41 | - | - | - | - |
| 2003-09-27 | 26 | 4.6 | 0.95 | - | 5.4 | 0.01 |
| 2004-06-05 | 43 | 3.9 | - | - | 2 | - |
| 2004-09-03 | 25 | 13 | 4.8 | - | 13 | 0.51 |
| 2004-10-07 | 35 | 18 | 3 | - | 12 | 1.2 |
| 2006-09-02 | 44.1 | 2.8 | - | - | - | - |
| 2008-07-23 | 35 | 0.36 | - | - | - | - |
| 2009-02-07 | 40 | 3.5 | - | - | 0.07 | - |
| 2009-09-04 | 28.3 | 2.3 | - | - | 1.5 | - |
| 2009-10-08 | 19.1 | 33 | 10 | 0.01 | 20 | 2.5 |
| 2009-11-21 | 38 | 18 | 1 | - | 11 | - |
| 2010-07-06 | 26 | 14 | 4.5 | - | 12 | 1.7 |
| 2010-09-03 | 33.3 | 3.8 | - | - | 6.8 | - |
| 2010-12-25 | 26 | 33 | 11 | - | 19 | 4.3 |
| 2013-04-21 | 40.7 | 2.5 | - | - | 1.8 | - |
| 2013-04-30 | 21.2 | 10 | 3.7 | 0.07 | 6.9 | 0.78 |
| 2013-10-12 | 22.2 | 3.5 | 1.4 | - | 2.9 | - |
| 2014-05-08 | 35.4 | 2.4 | - | - | - | - |
| 2014-08-23 | 22.2 | 7.6 | 5.7 | - | 13 | 0.49 |
| 2015-09-07 | 29.3 | 3.9 | 0.66 | - | 4.1 | - |
| 2016-02-06 | 31 | 13 | 2.5 | - | 11 | 0.62 |
| SUM | - | - | 55 | 0.09 | 155 | 13 |
| *2013-02-15* | *29.5* | *500* | - | - | *45* | *19* |
| TOTAL | - | - | - | - | 200 | 32 |